\documentclass[]{JHEP3}
\usepackage{epsfig}
\usepackage{cite}

\def\as{\alpha_{\mbox{\tiny S}}}
\def\rs{\rho_{\mbox{\tiny S}}}
\def\ee{e^+e^-}

\def\eps{\epsilon}

\def\0t{{\mbox{\bf 0}}}
\def\kt{{\mbox{\bf k}_\perp}}
\def\pt{{\mbox{\bf p}_\perp}}
\def\qt{{\mbox{\bf q}_\perp}}
\def\Qt{{\mbox{\bf Q}_\perp}}
\def\ptr#1{{\mbox{\bf p}_{\perp #1}}}
\def\qtr#1{{\mbox{\bf q}_{\perp #1}}}
\def\qb{{\bar q}}
\def\sb{{\bar s}}
\def\tb{{\bar t}}
\def\ub{{\bar u}}
\def\tk{\tilde{\kappa}}
\def\tq{\tilde{q}}
\def\frac#1#2{ {{#1} \over {#2} }}

\def\HW{{\small HERWIG}}
\def\HWP{{\small HERWIG++}}
\def\rat#1#2{\mbox{\small $\frac{#1}{#2}$}}
\def\half{\rat{1}{2}}

\def\beq{\begin{equation}}
\def\beeq{\begin{eqnarray}}
\def\eeq{\end{equation}}
\def\eeeq{\end{eqnarray}}

\title{New formalism for QCD parton showers}

\author{Stefan Gieseke$^{\dagger}$,  Philip Stephens$^{\dagger}$ 
and Bryan Webber$^{\dagger,\ddagger}$\\
$^{\dagger}$Cavendish Laboratory, University of Cambridge, Madingley Road, 
Cambridge, CB3~0HE, UK.\\
$^{\ddagger}$Theory Division, CERN, 1211 Geneva 23, Switzerland.}

\abstract{We present a new formalism for parton shower simulation
of QCD jets, which incorporates the following features: invariance
under boosts along jet axes, improved treatment of heavy quark fragmentation,
angular-ordered evolution with soft gluon coherence, more accurate soft gluon
angular distributions, and better coverage of phase space. It is implemented
in the new \HWP\ event generator.}


\preprint{Cavendish-HEP-03/18 \\
                CERN-TH/2003-239}

\begin{document}

\section{Introduction}
The parton shower approximation has become a key component of a wide
range of comparisons between theory and experiment in particle physics.
Calculations of infrared-safe observables, i.e.\ those
that are asymptotically insensitive to soft physics, can be
performed in fixed-order perturbation theory, but the resulting
final states consist of a few isolated partons, quite unlike the
multihadron final states observed experimentally. One can attempt
to identify isolated partons with hadronic jets, but then the
energy flows within and between jets are not well represented.

At present, the only means of connecting few-parton states with
the real world is via parton showers, which generate high-multiplicity
partonic final states in an approximation
that retains enhanced collinear and soft contributions to all orders.
Such multiparton states can be interfaced to a hadronization model
which does not require large momentum transfers in order to produce
a realistic hadronic final state.  Hadronization and detector
corrections to the fixed-order predictions can then be computed,
and the results have generally been found to be in satisfactory
agreement with the data.  Infrared-sensitive quantities such as hadron
spectra and multiplicities have also been described successfully
using parton showers. This increases confidence that similar techniques
can be used to predict new physics signals and backgrounds in
future experiments.

A crucial ingredient of modern parton showering algorithms\footnote{For
a general introduction to the parton shower approximation, see for
example Chapter 5 of \cite{Ellis:1996qj}.} is {\em angular ordering},
which ensures that important aspects of soft gluon coherence are included
in an azimuthally-averaged form.  The angular shower evolution 
variable \cite{Marchesini:1984bm} used in the event generator
program \HW\ \cite{Corcella:2001bw} is good for ensuring that angular
ordering is built in from the outset, but the phase space is complicated and
not invariant under any kind of boosts. Evolution in virtuality looks natural
but then angular ordering must be imposed afterwards, as is done
in {\small PYTHIA} \cite{Sjostrand:2000wi}.

In the present paper we investigate a new shower evolution formalism,
based on an angular variable related to transverse
momentum \cite{Catani:1991rr,Catani:1991hj,Catani:2000ef,Cacciari:2001cw}.
The main aim is to retain the direct angular ordering of the shower while
improving the Lorentz invariance of the evolution and simplifying the
coverage of phase space, especially in the soft region.  The new
shower variables also permit a better treatment of heavy quark
fragmentation, through evolution down to zero transverse momentum
and the use of mass-dependent splitting functions, which eliminate
the sharply-defined collinear ``dead cones'' seen in earlier
angular-ordered treatments.

In the following section we define the new shower variables and their
associated kinematics and dynamics, including the appropriate argument
of the running coupling, the mass-dependent parton branching
probability, and the shower evolution cutoff.  The variables are
defined slightly differently for initial- and final-state parton
branching, and depend on the colour connection of the evolving parton,
so we consider in subsequent sections the various possible
configurations of colour flow between initial and final jets.

The formalism presented here is implemented in the new Monte Carlo event
generator \HWP\ \cite{Gieseke:2003_20}. Results for $e^+e^-$ annihilation
and comparisons with LEP data will be presented shortly in a separate
publication~\cite{Gieseke:2003_19}.
The formulae in the present paper could also be used to construct
a matching scheme for next-to-leading order (NLO) QCD calculations
and \HWP\ parton showers, similar to that developed for \HW\ showers
in \cite{Frixione:2002ik,Frixione:2003ei} and implemented in
the {\small MC@NLO} event generator~\cite{Frixione:2003vm},
or to improve schemes for combining parton showers with fixed-order
matrix elements~\cite{Catani:2001cc}.

\section{New variables for parton branching} 
\subsection{Final-state quark branching}
\subsubsection{Kinematics}
\begin{figure}[htb]
%\vspace{9pt}
\begin{center}
\epsfig{figure=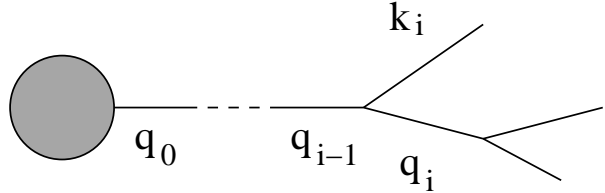,width=8cm}
\end{center}
\caption{Final-state parton branching. The blob represents the
hard subprocess.}
\label{fig_finalbr}
\end{figure}

Consider parton branching in an outgoing (heavy) quark jet. Define the quark
momentum after the $i$th gluon emission $q_{i-1}\to q_i+k_i$ (see
fig.~\ref{fig_finalbr}) as
\beq\label{eq_qi}
q_i = \alpha_i p + \beta_i n + q_{\perp i}
\eeq
where $p$ is the jet's ``parent parton'' momentum ($p^2=m^2$,
the {\em on-shell} quark mass-squared),
$n$ is a lightlike ``backward'' 4-vector ($n^2=0$), and $q_{\perp i}$
is the transverse momentum ($q_{\perp i}^2=-\qt_i^2$, 
$q_{\perp i}\cdot p = q_{\perp i}\cdot n =0$).  Then
\beq\label{eq_betai}
\beta_i = \frac{\qt_i^2 + q_i^2 - \alpha_i^2 m^2}{2\alpha_i p\cdot n}\;.
\eeq

The momentum fraction and relative transverse momentum are now defined as
\beq\label{eq_zipti}
z_i = \frac{\alpha_i}{\alpha_{i-1}}\;,\;\;\;\;\;\;
\ptr{i} = \qtr{i} - z_i\mbox{\bf q}_{\perp i-1}\;.
\eeq
Then we have
\beq\label{eq_qi-1}
q_{i-1}^2 = \frac{q_i^2}{z_i}+\frac{k_i^2}{1-z_i}+
   \frac{\ptr{i}^2}{z_i(1-z_i)}\;.
\eeq

\subsubsection{Running coupling}
To find the optimal argument of $\as$, we consider the branching of a
quark of virtuality $q^2$ into an on-shell quark and an off-shell gluon
of virtuality $k^2$ \cite{Dokshitzer:1996qm}. From eq.~(\ref{eq_qi-1}),
the propagator denominator is
\beq\label{eq_q2m2}
q^2-m^2 = \frac{1-z}{z}m^2 +\frac{k^2}{1-z}+ \frac{\pt^2}{z(1-z)}
 = \frac{1}{1-z}\left\{k^2 +\frac{1}{z}[\pt^2 + (1-z)^2 m^2]\right\}\;.
\eeq
The dispersion relation for the running coupling is supposed to be
\beq
\frac{\as(\mu^2)}{\mu^2} = \frac{\as(0)}{\mu^2}
+ \int_0^\infty \frac{dk^2}{k^2(k^2+\mu^2)}\rs(k^2)
\eeq
where $\rs(k^2)$ is the discontinuity of $\as(-k^2)$.  The first term
on the right-hand side comes from cutting through the on-shell gluon,
the second from cutting through the gluon self-energy. In our case we
have $k^2 + [\pt^2 + (1-z)^2 m^2]/z$ in place of $k^2+\mu^2$. We are interested
in soft gluon resummation ($z\to 1$) \cite{Cacciari:2001cw} and so we
ignore the factor of $1/z$ here. Thus the suggested argument of $\as$ is
$\pt^2 + (1-z)^2 m^2$.  In practice we impose a minimum virtuality on
light quarks and gluons in the parton shower, and therefore the actual
argument is slightly more complicated (see below).

\subsubsection{Evolution variable}
The evolution variable is not simply $q^2$ since this would ignore
angular ordering. For massless parton branching this
means the evolution variable should be $\pt^2/[z(1-z)]^2= q^2/z(1-z)$
\cite{Marchesini:1983bm}. For gluon emission by a massive quark we assume
this generalizes to $(q^2-m^2)/z(1-z)$. To define a resolvable emission
we also need to introduce a minimum virtuality $Q_g^2$ for gluons and
light quarks.  Therefore from
eq.~(\ref{eq_q2m2}) the evolution variable is
\beq\label{eq_ftq}
\tq^2 = \frac{\pt^2}{z^2(1-z)^2} + \frac{\mu^2}{z^2}
 + \frac{Q_g^2}{z(1-z)^2}
\eeq
where $\mu = \max(m,Q_g)$.  For the argument of the running coupling
we use
\beq\label{eq_asarg}
z^2(1-z)^2\tq^2 = \pt^2 + (1-z)^2\mu^2 + z Q_g^2\;.
\eeq
Note that for massive quarks this allows us to evolve down to
$\pt<(1-z)m$, i.e.\ inside the
{\em dead cone} \cite{Marchesini:1990yk,Dokshitzer:fd}.

Angular ordering of the branching $q_i\to q_{i+1}$ is defined by
\beq
\tq_{i+1} < z_i\tq_i\;.
\eeq
The factor of $z_i$ enters because the angle at each branching is inversely
proportional to the momentum fraction of the parent.
Similarly for branching on the gluon, $k_i\to k_{i+1}$, we require
\beq
\tilde k_{i+1} < (1-z_i)\tq_i\;.
\eeq

\subsubsection{Branching probability}
For the parton branching probability we use the mass-dependent
splitting functions of ref.~\cite{Catani:2000ef}.  These are
derived in the {\em quasi-collinear limit}, in which $\pt^2$
and $m^2$ are treated as small (compared to $p\cdot n$) but
$\pt^2/m^2$ is not necessarily small.  In this limit the $q\to qg$
splitting function is
\beq
P_{qq}(z,\pt^2) = C_F\left[\frac{1+z^2}{1-z}
-\frac{2z(1-z)m^2}{\pt^2+(1-z)^2m^2}\right]\;.
\eeq
Note that at $\pt=0$ the factor in square brackets is just
$1-z$, i.e.\ the soft singularity at $z\to 1$ becomes a zero
in the collinear direction. The minimum virtuality $Q_g^2$ serves
only to define a resolvable emission, and therefore we omit it
when defining the branching probability in terms of the evolution
variable (\ref{eq_ftq}) as
\beq\label{eq_dPqqg}
dP(q\to qg) = \frac{\as}{2\pi}\frac{d\tq^2}{\tq^2}\,P_{qq}\,dz
= \frac{C_F}{2\pi}\as[z^2(1-z)^2\tq^2]\frac{d\tq^2}{\tq^2}\frac{dz}{1-z}
\left[1+z^2-\frac{2m^2}{z\tq^2}\right]\;.
\eeq

\subsection{Gluon splitting}
\label{sec:gsplit}
In the case of a final-state gluon splitting into a pair of heavy quarks of
mass $m$, the quasi-collinear splitting function derived
in \cite{Catani:2000ef} is
\beq\label{eq:gQQ2}
P_{qg}(z,\pt^2) = 
  T_R\left[1-2z(1-z)\frac{\pt^2}{\pt^2 + m^2}\right]\,. 
\eeq
We note that this splitting function is bounded above by its value $T_R=\half$
at the phase space boundary $\pt=\0t$, and below by $T_R/2$.
By analogy with eq.~(\ref{eq_ftq}), in this case the evolution
variable $\tq$ is related to the virtuality of the gluon or the relative
transverse momentum of the splitting by
\beq\label{eq:qtgqq}
  \tq^2 = \frac{q^2}{z(1-z)} = \frac{\pt^2+m^2}{z^2(1-z)^2}\,.
\eeq
In terms of the variables $\tilde q, z$, the $g\to q\qb$
branching probability then reads
\beq\label{eq:gQQ}
dP(g\to q\qb) = \frac{T_R}{2\pi}\as[z^2(1-z)^2\tq^2]\frac{d\tq^2}{\tq^2}
\left[1-2z(1-z)+\frac{2m^2}{z(1-z)\tq^2}\right]\,dz\,. 
\eeq

In the case of gluon splitting into gluons, the branching probability
takes the familiar form
\beq\label{eq:ggg}
dP(g\to gg) = \frac{C_A}{2\pi}\as[z^2(1-z)^2\tq^2]\frac{d\tq^2}{\tq^2}
\left[\frac{z}{1-z}+\frac{1-z}{z}+z(1-z)\right]\,dz\,. 
\eeq
Since we introduce a minimum virtuality $Q_g^2$ for gluons,
the relationship between the evolution variable and the relative
transverse momentum for this splitting is as in eq.~(\ref{eq:qtgqq}) but
with the heavy quark mass $m$ replaced by $Q_g$. Similarly, for gluon
splitting to light quarks we use eq.~(\ref{eq:qtgqq}) with
$\mu = \max(m,Q_g)$ in place of $m$.

\subsection{Initial-state branching}
\begin{figure}[htb]
%\vspace{9pt}
\begin{center}
\epsfig{figure=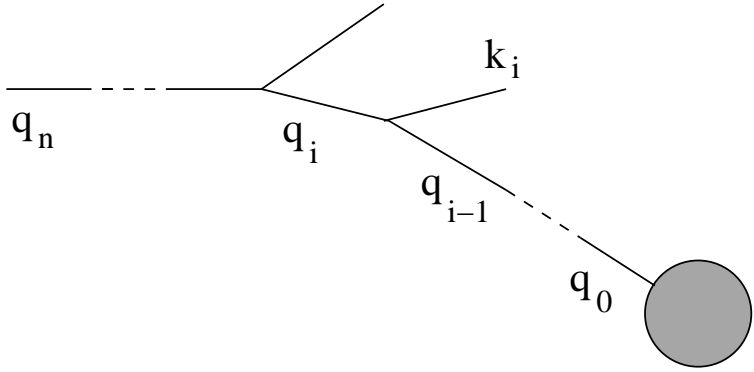,width=10cm}
\end{center}
\caption{Initial-state parton branching. The blob represents the
hard subprocess.}
\label{fig_initbr}
\end{figure}

Consider the initial-state (spacelike) branching of a partonic
constituent of an incoming hadron that undergoes some hard
collisions subprocess such as deep inelastic lepton scattering.
The momenta are defined as in eq.~(\ref{eq_qi}), with the reference
vector $p$ along the beam direction.
In this case the evolution is performed {\em backwards} from the hard
sub-process to the incoming hadron, as shown in fig.~\ref{fig_initbr}.
Thus we now define in place of eq.~(\ref{eq_zipti})
\beq\label{eq_zispa}
z_i = \frac{\alpha_{i-1}}{\alpha_i}\;,\;\;\;\;\;\;
\ptr{i} = \qtr{i-1} - z_i\qtr{i}\;.
\eeq
Then
\beq\label{eq_qispa}
q_{i-1}^2 = z_i q_i^2 -\frac{z_i}{1-z_i}k_i^2 - \frac{\ptr{i}^2}{1-z_i}\;.
\eeq

We assume a massless variable-flavour-number evolution
scheme~\cite{Aivazis:1993pi,Thorne:1997ga} for constituent
parton branching, setting $m=0$ and putting all emitted
gluons at the minimum virtuality, $k_i^2=Q_g^2$.  The angular evolution
variable now relates only to the angle of the emitted gluon
and therefore we choose
\beq\label{eq_itqi}
\tq_i^2 =  \frac{\ptr{i}^2 +z_i Q_g^2}{(1-z_i)^2}\;,
\eeq
with ordering condition simply $\tq_{i+1} < \tq_i$.
Correspondingly, for the argument of the running coupling we now use
$(1-z)^2\tq^2$.

A different type of initial-state branching occurs in the decay of
heavy, quasi-stable coloured objects like the top quark.  Here the
momentum of the incoming heavy object is fixed and evolution is
performed forwards to the hard decay process. In this case we
cannot neglect the mass of the parton and eq.~(\ref{eq_itqi})
becomes
\beq\label{eq_dec_tqi}
\tq_i^2 =  \frac{\ptr{i}^2+z_i Q_g^2}{(1-z_i)^2}+m^2\;,
\eeq
while the branching probability (\ref{eq_dPqqg}) is replaced by
\beq\label{eq_dec_dP}
dP(q\to qg) = \frac{C_F}{2\pi}\as[(1-z)^2\tq^2]
\frac{d\tq^2}{\tq^2}\frac{dz}{1-z}
\left[1+z^2-\frac{2zm^2}{\tq^2}\right]\;.
\eeq

\subsection{Allowed regions and termination of branching}
The allowed phase space for each branching is given by requiring a real
relative transverse momentum, $\pt^2> 0$.  In final-state $q\to qg$
branching, we have from eq.~(\ref{eq_asarg})
\beq\label{eq_qqg_ps}
z^2(1-z)^2\tq^2 > (1-z)^2\mu^2 + z Q_g^2\;.
\eeq
This yields a rather complicated boundary in the $(\tq,z)$ plane.
However, since
\beq
(1-z)^2\mu^2 + z Q_g^2 > (1-z)^2\mu^2\,,\; z^2 Q_g^2
\eeq
we see that the phase space lies inside the region
\beq
  \label{eq:qtzlimits}
  \frac{m}{\tq} < z < 1-\frac{Q_g}{\tq}\;, 
\eeq
and approaches these limits for large values of $\tq$.  The precise
phase space can therefore be filled efficiently by generating
values of $z$  between these limits and rejecting those that violate
the inequality (\ref{eq_qqg_ps}).
The resulting threshold for $\tq$ is slightly larger than but of
the order of $m+Q_g$.

In gluon splitting, we obtain the allowed phase space range
from eq.~(\ref{eq:qtgqq}) as
\beq\label{eq:zrangegqq}
  z_- < z < z_+, \qquad 
  z_{\pm} = \frac{1}{2}\left(1\pm \sqrt{1-\frac{4\mu}{\tq}}\right) 
  \qquad\mathrm{ and }\quad
  \tq > 4\mu
\eeq
where $\mu=m$ for splitting into heavy quarks, or $\mu=\max(m,Q_g)$
more generally.  Therefore, analogously to eq.~(\ref{eq:qtzlimits}),
the phase space lies within the range
\beq
  \frac{\mu}{\tq} < z < 1-\frac{\mu}{\tq}\;. 
\eeq

Schematically, the parton shower corresponds to selecting a sequence of
$(\tq_i,z_i)$ values by solving the equations
\beeq
R_1 &=& \exp\left(-\int_{\tq_i}^{\tq_{i-1}}d\tq\int_{z_-}^{z_+}dz\,
\frac{d^2P}{d\tq\,dz}\right)\nonumber \\
R_2 &=& \int_{z_-}^{z_i}dz\,\frac{d^2P}{d\tq\,dz}\Bigg/\int_{z_-}^{z_+}dz\,
\frac{d^2P}{d\tq\,dz}
\eeeq
where $R_{1,2}\in [0,1]$ are uniform pseudorandom numbers.
Whenever the algorithm selects a value of $\tq$ below
the threshold, branching of that parton is terminated.
The minimum virtuality $Q_g$ thus determines the scale at which soft or
collinear parton emission becomes unresolvable.  In the absence
of such a scale one eventually reaches a region where the perturbative
expression for the running coupling is divergent.

One may wish to use a parametrization of $\as$ at low scales such that
$\as(0)$ is finite.  However, a cutoff $Q_g$ is still needed
to avoid divergence of the $q\to qg$ and $g\to gg$ branching probabilities.
Alternatively one could consider parametrizing $\as$ such that
$\as(0)=0$, e.g.
\beq
\as(q^2) = \frac{q}{Q_c}\as(Q_c^2)\;\;\;\;\mbox{for}\; q<Q_c\;,
\eeq
where $Q_c>\Lambda$.  Then the total branching probability below
$Q_c$ is (for massless quarks)
\beq
P_c(q\to qg) = C_F\frac{\as(Q_c^2)}{2\pi}\int_0^1 2z(1+z^2)\,dz
=\frac{\as(Q_c^2)}{\pi}\;,
\eeq
and no explicit cutoff is required, although of course $Q_c$ is
essentially playing the same r\^ole.

After branching has terminated, the outgoing partons are put on
mass-shell (or given the virtual mass $Q_g$ if lighter) and the
relative transverse momenta of the branchings in the shower
are computed.  For final-state gluon splitting we have
\beq
|\pt| = \sqrt{z^2(1-z)^2 \tq^2 - \mu^2}, 
\eeq
or else, if the parent is a quark, 
\beq
|\pt| = \sqrt{ (1-z)^2 (z^2 \tq^2 - \mu^2) - z Q_g^2 }.
\eeq
The virtualities of the internal lines of the shower can now be
computed backwards according to eq.~(\ref{eq_qi-1}). Finally, the azimuthal
directions of the $\pt$'s can be chosen \cite{Richardson:2001df} and the
full 4-momenta reconstructed using eqs.~(\ref{eq_qi}) and (\ref{eq_betai}).

In initial-state constituent parton branching the evolution is ``guided''
by the parton distribution functions (PDFs) of the incoming parent hadron.
Since PDFs are often not tabulated below some scale $Q_s>Q_0$, one may wish
to terminate branching whenever $\tq <Q_s$ is selected. In that case
the incoming parton is assigned virtuality $q_n^2\sim -Q_s^2$ and
the spacelike virtualities of internal lines are then reconstructed
back from $q_n^2$ to $q_0^2$ using the transverse momenta deduced
from eq.~(\ref{eq_itqi}) inserted in  eq.~(\ref{eq_qispa}).

For initial-state branching in the decay of a heavy, quasi-stable
coloured object, the branching proceeds in the opposite direction but
the reconstruction of momenta is similar, using eq.~(\ref{eq_dec_tqi})
instead of (\ref{eq_itqi}).

\subsection{Treatment of colour flows}
The more detailed treatment depend on the choice of the ``backward'' vector
$n$ and on which quantities are to be held fixed during jet evolution.
Normally $n$ should be taken along the colour-connected partner of the
radiating parton, and the 4-momentum of the colour-connected system
should be preserved. The upper limits on the evolution variable $\tq$
for the  colour-connected jets should be chosen so as to cover the phase
space in the soft limit, with the best possible approximation to the
correct angular distribution. In setting these limits we neglect the
minimum virtuality $Q_g^2$, which is a good approximation at high
energies. We consider separately the four cases that
the colour connection is between two final-state jets, two initial-state
(beam) jets, a beam jet and a final-state jet, or a decaying heavy parton
and a decay-product jet.

\section{Final-final colour connection}\label{sec_finfin}
Consider the process $a\to b+c$ where $a$ is a colour singlet and $b$ and
$c$ are colour-connected.  Examples are $\ee\to q\qb$ and $W\to q\qb'$.
We need to preserve the 4-momentum of $a$ and therefore we work
in its rest-frame,
\beq
p_a = Q(1,\0t,0)\;,\;\;\; p_b= \half Q(1+b-c,\0t,\lambda)
\;,\;\;\; p_c= \half Q(1-b+c,\0t,-\lambda)\;,
\eeq
where $p_a^2=Q^2$, $b=m_b^2/Q^2$, $c=m_c^2/Q^2$ and
\beq
\lambda = \lambda(1,b,c)\equiv\sqrt{(1+b-c)^2-4b} = \sqrt{(1-b+c)^2-4c}\;.
\eeq
For emission of a gluon $g$ from $b$ we write
\beq\label{eq_ffqi}
q_i = \alpha_i p_b + \beta_i n + q_{\perp i}
\eeq
where $\qtr{g}=\kt$, $\qtr{b}=-\kt$, $\qtr{c}=\0t$ and we choose
\beq
n = \half Q(\lambda,\0t,-\lambda)\;.
\eeq
Notice that, if $c$ is massive, the alignment of $n$ along $p_c$ is
exact only in a certain class of Lorentz frames.  However, if we try to
use a massive ``backward'' vector the kinematics become too complicated.

To preserve $p_a=q_b+q_c+q_g$ we require
\beq\label{eq_ffsums}
\sum\alpha_i = \sum\beta_i = \frac{2}{1+b-c+\lambda}
\eeq
whereas the mass-shell conditions give
\beeq\label{eq_ffbetas}
\beta_b &=& \frac{2}{\lambda(1+b-c+\lambda)}
\left(\frac{b+\kappa}{\alpha_b}-b\alpha_b\right) \nonumber \\
\beta_c &=& \frac{2}{\lambda(1+b-c+\lambda)}
\left(\frac{c}{\alpha_c}-b\alpha_c\right) \\
\beta_g &=& \frac{2}{\lambda(1+b-c+\lambda)}
\left(\frac{\kappa}{\alpha_g}-b\alpha_g\right) \nonumber
\eeeq
where $\kappa\equiv\kt^2/Q^2$.  Our new variables are
\beq\label{eq_fztk}
z = \frac{\alpha_b}{\alpha_b+\alpha_g}\;,\;\;\;
\tk \equiv \frac{\tq^2}{Q^2} > b\;,
\eeq
where from eq.~(\ref{eq_ftq}) we have
\beq\label{eq_fkt}
\kappa= (z^2\tk-b)(1-z)^2\;,
\eeq
and so $\sqrt{b/\tk}<z<1$. From eqs.(\ref{eq_ffsums})-(\ref{eq_fkt}) we find
\beeq\label{eq_ffalphas}
\alpha_b &=&\frac{z}{1+b-c+\lambda}\left(1+b-c+ z(1-z)\tk+
\sqrt{[1-b+c- z(1-z)\tk]^2-4b}\right)\;,\nonumber \\
\alpha_c &=& \frac{2}{1+b-c+\lambda}-\frac{\alpha_b}{z}\;,\\
\alpha_g &=&\frac{1-z}{z}\alpha_b\;,\nonumber
\eeeq
with the $\beta_i$'s given by eq.~(\ref{eq_ffbetas}).

\subsection{Phase space variables}
It is convenient to express the phase space in terms of the Dalitz plot
variables
\beq
x_i = \frac{2p_a\cdot q_i}{Q^2} = (1+b-c)\alpha_i + \lambda\beta_i\;.
\eeq
Substituting from eqs.~(\ref{eq_ffbetas}) and (\ref{eq_ffalphas}), we find
\beeq\label{eq_xcbg}
x_c &=& 1-b+c- z(1-z)\tk \nonumber \\
x_b &=& (2-x_c)r + (z-r)\sqrt{x_c^2-4c}\\
x_g &=& (2-x_c)(1-r) - (z-r)\sqrt{x_c^2-4c}\nonumber
\eeeq
where
\beq\label{eq_r}
r = \frac 12\left(1+\frac{b}{1+c-x_c}\right)\;.
\eeq
The Jacobian factor is thus simply
\beq
\frac{\partial(x_b,x_c)}{\partial(z,\tk)} = z(1-z)\sqrt{x_c^2-4c}
\eeq
and the quasi-collinear branching probability (\ref{eq_dPqqg})
translates to
\beq\label{eq_dPx}
dP(q\to qg) = C_F\frac{\as}{2\pi}\frac{dx_b\,dx_c}{(1-b+c-x_c)\sqrt{x_c^2-4c}}
\left[\frac{1+z^2}{1-z}-\frac{2b}{1-b+c-x_c}\right]
\eeq
where
\beq
z= r +\frac{x_b-(2-x_c)r}{\sqrt{x_c^2-4c}}\;,
\eeq
$r$ being the function of $x_c$ given in eq.~(\ref{eq_r}).

For emission from parton $c$ we write
\beq
q_i = \alpha_i p_c + \beta_i n + q_{\perp i}
\eeq
where now we choose
\beq
n = \half Q(\lambda,\0t,\lambda)\;.
\eeq
Clearly, the region covered and the branching probability
will be as for emission from parton $b$, but with $x_b$ and
$x_c$, $b$ and $c$ interchanged.

\subsection{Soft gluon region}\label{sec_finfin_sof}
For emission from parton $b$ in the soft region $1-z=\eps\to 0$ we have
\beq
x_c\sim 1-b+c-\eps\tk\;,\;\;\;
x_b\sim 1+b-c-\eps\tk'
\eeq
where
\beq\label{eq_soft}
\tk' = \lambda+\frac{\tk}{2b}(1-b-c-\lambda)\;.
\eeq
Since $\tk$ is an angular variable, we can express it in terms of
the angle $\theta_{bg}$ between the directions of the emitting
parton $b$ and the emitted gluon in the rest frame of $a$.
In the soft region we find
\beq
\tk = \frac{(1+b-c+\lambda)(1+b-c-\lambda\cos\theta_{bg})}
{2(1+\cos\theta_{bg})}
\eeq
Thus $\tk=b$ at $\theta_{bg}=0$ and $\tk\to\infty$ as $\theta_{bg}\to\pi$. 

For soft emission from parton $c$, the roles of $x_b$ and $x_c$,
$b$ and $c$ are interchanged. To cover the whole angular region
in the soft limit, we therefore require $\tk<\tk_b$ in jet $b$ and
 $\tk<\tk_c$ in jet $c$, where
\beq
\frac{\tk_b}{\tk'_b} = \frac{\tk'_c}{\tk_c}
\eeq
and hence
\beq\label{eq_fftkbc}
(\tk_b -b)(\tk_c -c) =  \frac 14 (1-b-c+\lambda)^2\;.
\eeq
In particular, the most symmetric choice is
\beq\label{eq_fftks}
\tk_b = \half(1+b-c+\lambda)\;,\;\;\;
\tk_c = \half(1-b+c+\lambda)\;.
\eeq
The largest region that can be covered by one jet corresponds to the maximal
value of $\tk$ allowed in eq.~(\ref{eq_ffalphas}) for real $\alpha_b$,
i.e.\ for the maximal $b$ jet
\beq\label{eq_fftkmax}
\tk_b = 4(1-2\sqrt b -b+c)\;.
\eeq

\subsection{Example: $\ee\to q\bar q g$}
\begin{figure}[htb]
%\vspace{9pt}
\begin{center}
\epsfig{figure=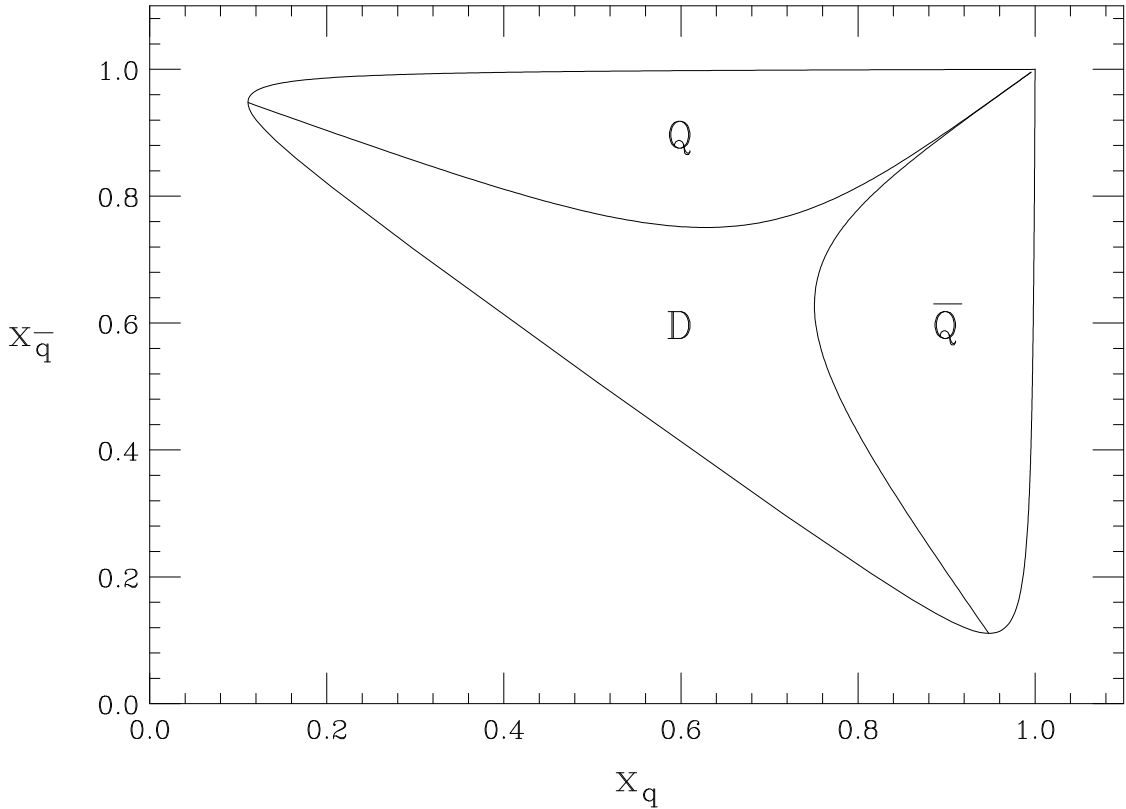,angle=90,width=15cm}
\end{center}
\caption{Phase space for $e^+e^-\to q\qb g$ for $m_q=5$ GeV, $Q^2=m_Z^2$,
with symmetric definition of quark and antiquark jets.}
\label{fig_bfragn}
\end{figure}

Here we have $b=c=\rho$, $\lambda = \sqrt{1-4\rho}=v$, the quark
velocity in the Born process $\ee\to q\bar q$.  The phase space and
the two jet regions for the symmetrical choice (\ref{eq_fftks}) are
shown in fig.~\ref{fig_bfragn}. The region D, corresponding to hard
non-collinear gluon emission, is not included in either jet and must
be filled using the ${\cal O}(\as)$ matrix element (see below).

\begin{figure}[htb]
%\vspace{9pt}
\begin{center}
\epsfig{figure=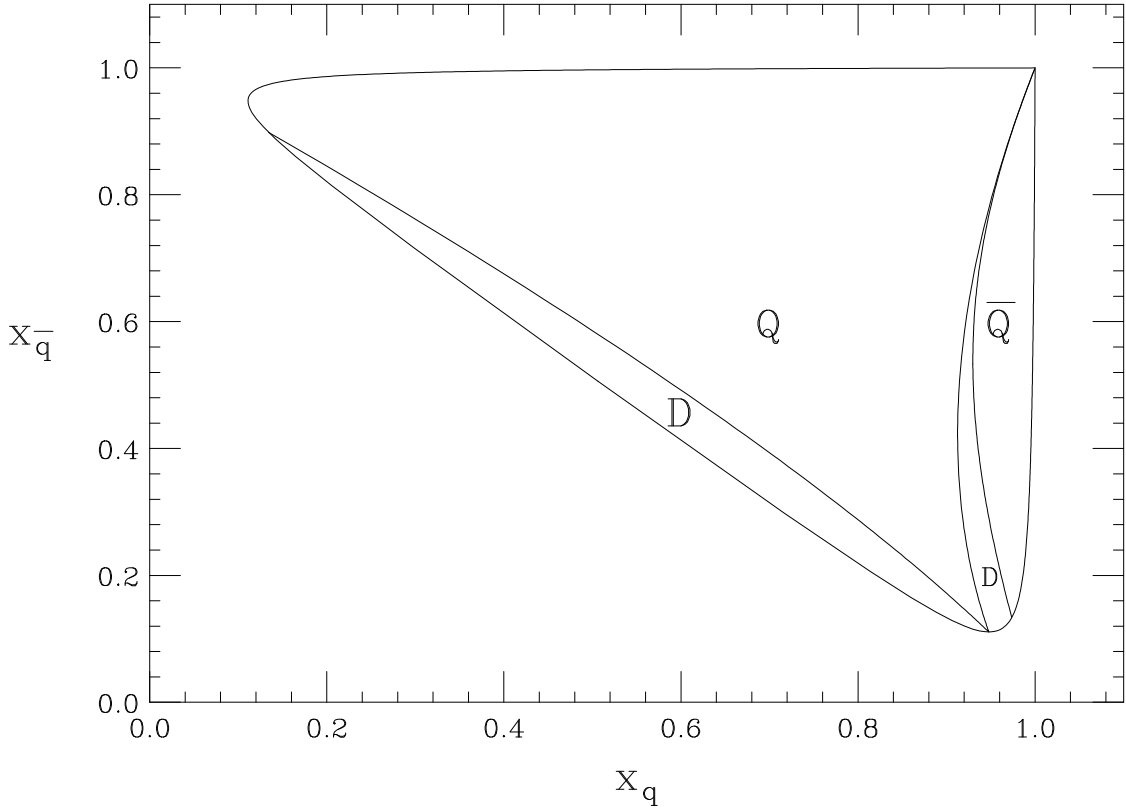,angle=90,width=15cm}
\end{center}
\caption{Phase space for $e^+e^-\to q\qb g$ for $m_q=5$ GeV, $Q^2=m_Z^2$,
with maximal region for the quark jet.}
\label{fig_bfragn_max}
\end{figure}

For the maximal quark jet we get from eq.~(\ref{eq_fftkmax})
\beq
\tk_q = 4(1-2\sqrt\rho)\;,
\eeq
as shown in fig.~\ref{fig_bfragn_max} together with the complementary
antiquark jet region given by eq.~(\ref{eq_fftkbc}).

\subsubsection{Exact matrix element}
The $\ee\to V\to q\qb g$ differential cross section,
where $V$ represents a vector current such as a virtual photon,
is given to first order in $\as$ by \cite{Nason:1994xx,Nason:1997pk}
\beq\label{eq_dsigdxdxb}
\frac{1}{\sigma_V}\frac{d^2\sigma_V}{dx_q\,dx_\qb} =
\frac{\as}{2\pi}\frac{C_F}{v}
\left[ \frac{(x_q+2\rho)^2+(x_\qb+2\rho)^2+\zeta_V}
{(1+2\rho)(1-x_q)(1-x_\qb)} -\frac{2\rho}{(1-x_q)^2}
-\frac{2\rho}{(1-x_\qb)^2}\right]
\eeq
where
\beq\label{eq_zetaV}
\zeta_V = -8\rho(1+2\rho)
\eeq
and
\beq\label{eq_sigmaQ}
\sigma_V = \sigma_0\left(1+2\rho\right)v
\eeq
is the Born cross section for heavy quark production by a vector
current, $\sigma_0$ being the massless quark Born cross section.

In the case of the axial current contribution $\ee\to A\to q\qb g$,
instead of eq.~(\ref{eq_dsigdxdxb}) we have
\beq\label{eq_dsigA}
\frac{1}{\sigma_A}\frac{d^2\sigma_A}{dx_q\,dx_\qb} =
\frac{\as}{2\pi}\frac{C_F}{v}
\left[\frac{(x_q+2\rho)^2+(x_\qb+2\rho)^2+\zeta_A}
{v^2(1-x_q)(1-x_\qb)}
-\frac{2\rho}{(1-x_q)^2}
-\frac{2\rho}{(1-x_\qb)^2}\right]\;,
\eeq
where
\beq\label{eq_zetaA}
\zeta_A = 2\rho[(3+x_g)^2-19+4\rho]\;,
\eeq
$\sigma_A$ being the Born cross section for heavy quark
production by the axial current:
\beq\label{eq_sigmaA}
\sigma_A = \sigma_0v^3\;.
\eeq

\subsubsection{Soft gluon distribution}
In the soft gluon region $1-z=\eps\to 0$ the branching probability
(\ref{eq_dPx}) becomes
\beeq\label{eq_dPsoft}
\frac{d^2P}{dx_q\,dx_\qb} &\sim& \frac{\as}{2\pi}\frac{2C_F}{v\eps^2}f_s(\tk)
\nonumber\\
f_s(\tk)&=&\frac{1}{\tk} -\frac{\rho}{\tk^2}\;.
\eeeq

\begin{figure}[htb]
%\vspace{9pt}
\begin{center}
\epsfig{figure=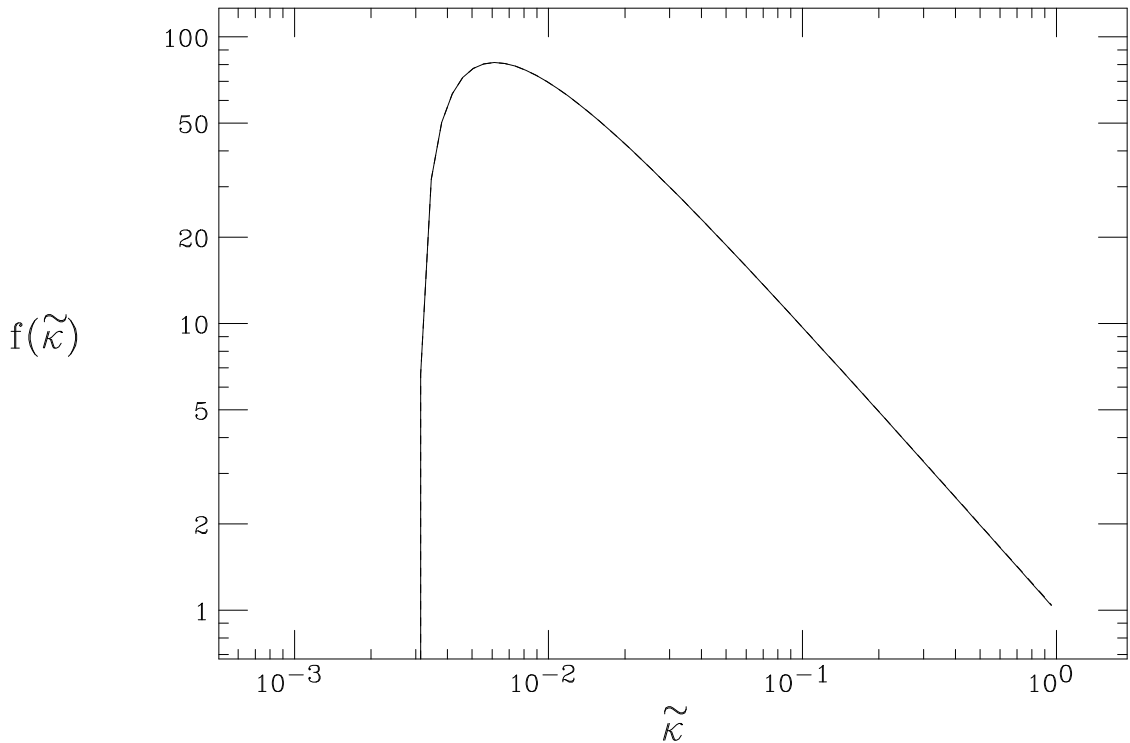,angle=90,width=15cm}
\end{center}
\caption{The function $f(\tk)$ giving the gluon angular distribution in
the soft limit, for $m=5$ GeV, $Q^2=m_Z^2$. The exact result
eq.~({\protect\ref{eq_dPVsoft}}), solid curve, and shower approximation
({\protect\ref{eq_dPsoft}}), dashed, are not distinguishable on this scale.}
\label{fig_soft}
\end{figure}

In this limit, the exact vector and axial current matrix elements,
eqs.~(\ref{eq_dsigdxdxb}) and (\ref{eq_dsigA}) respectively, give
identical distributions:
\beeq\label{eq_dPVsoft}
\frac{1}{\sigma_V}\frac{d^2\sigma_V}{dx_q\,dx_\qb}&\sim&
\frac{1}{\sigma_A}\frac{d^2\sigma_A}{dx_q\,dx_\qb}
\sim \frac{\as}{2\pi}\frac{2C_F}{v\eps^2}f(\tk)
\nonumber\\
f(\tk)&=&\frac{1-2\rho}{\tk\tk'}
-\frac{\rho}{\tk^2}-\frac{\rho}{\tk'^2}
\nonumber\\
&=&f_s(\tk)\left(\frac{v}{\tk'}\right)^2\;.
\eeeq
Since from eq.~(\ref{eq_soft})
\beq
\tk' = v+\tk\left(\frac{1-v}{1+v}\right) > v\;,
\eeq
the parton shower approximation
(\ref{eq_dPsoft}) always overestimates the true result in the soft limit,
and so correction by the rejection method is straightforward.
For small values of $\rho$ we have
\beq\label{eq_dPVlorho}
f(\tk)= \frac{1}{\tk}-\frac{\rho}{\tk^2}
+\frac{2\rho^2}{\tk}-2\rho+{\cal O}(\rho^2)\;.
\eeq
Since $\tk>\rho$ we see that the error in the approximation
(\ref{eq_dPsoft}) is at most ${\cal O}(\rho)$, for any value of $\tk$
(fig.~\ref{fig_soft}).

\subsubsection{Dead region contribution}
\begin{figure}[htb]
%\vspace{9pt}
\begin{center}
\epsfig{figure=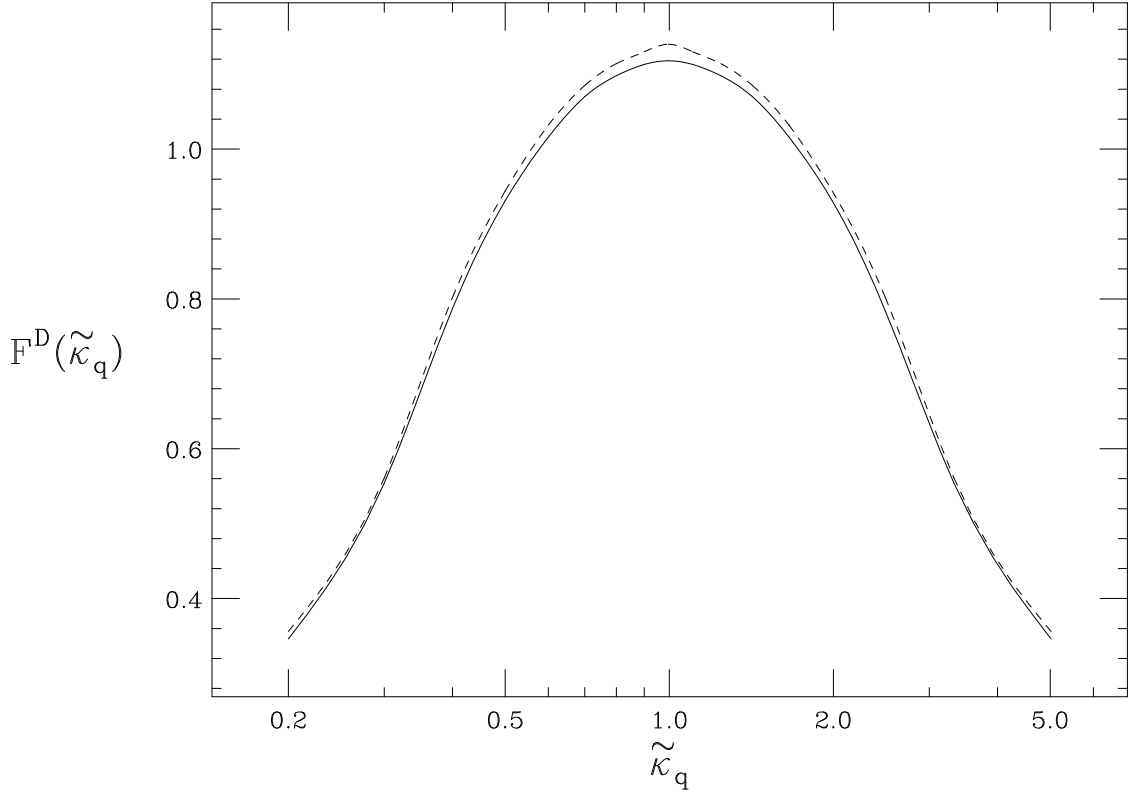,angle=90,width=15cm}
\end{center}
\caption{The function $F^D(\tk_q)$ giving the contribution of the dead
region to the cross section, for $m=5$ GeV, $Q^2=m_Z^2$.
Solid: vector current. Dashed: axial current.}
\label{fig_fdead}
\end{figure}

The integral over the dead region may be expressed as
\beq\label{eq_dsigdead}
\frac{1}{\sigma_V}\int_D d^2\sigma_V \equiv
\frac{\as}{2\pi} C_F\, F^D_V(\tk_q)
\eeq
where $\tk_q$ parametrizes the boundary of the quark jet.  As shown in
fig.~\ref{fig_fdead}, this is actually maximal, but still small,
at the symmetric point given by eq.~(\ref{eq_fftks}).
%However, the important issue in choosing the initial conditions for the
%parton showers is to minimize the overall discrepancy between the exact
%matrix element and the parton shower approximation (\ref{eq_dPx}),
%which is characterised by the difference of integrals
%\beq\label{eq_dsigtot}
%\frac{1}{\sigma_V}\int_{Q+\bar Q+D} d^2\sigma_V -\int_{Q+\bar Q} d^2P \equiv
%\frac{\as}{2\pi} C_F\, F_V(\tk_q)\;.
%\eeq
%Owing to the small ${\cal O}(\rho)$ discrepancy in the soft gluon limit
%(see above) we have to impose a cutoff $x_g>\eps$ in eq.~(\ref{eq_dsigtot}).
%Figure \ref{fig_ftot} shows the result for $\eps=10^{-4}$.
%We see that the overall discrepancy is negative (since the shower
%approximation overestimates the matrix element in the jet regions), and its
%magnitude is {\em minimal} at the symmetric point.
%\begin{figure}[htb]
%%\vspace{9pt}
%\begin{center}
%\epsfig{figure=ftot.ps,angle=90,width=15cm}
%\end{center}
%\caption{The function $F(\tk_q)$ giving the overall discrepancy between the
%matrix element and the parton shower approximation at ${\cal O}(\as)$,
%for $m=5$ GeV, $Q^2=m_Z^2$, with cutoff $x_g>10^{-4}$.
%Solid: vector current. Dashed: axial current.}
%\label{fig_ftot}
%\end{figure}

\begin{figure}[htb]
%\vspace{9pt}
\begin{center}
\epsfig{figure=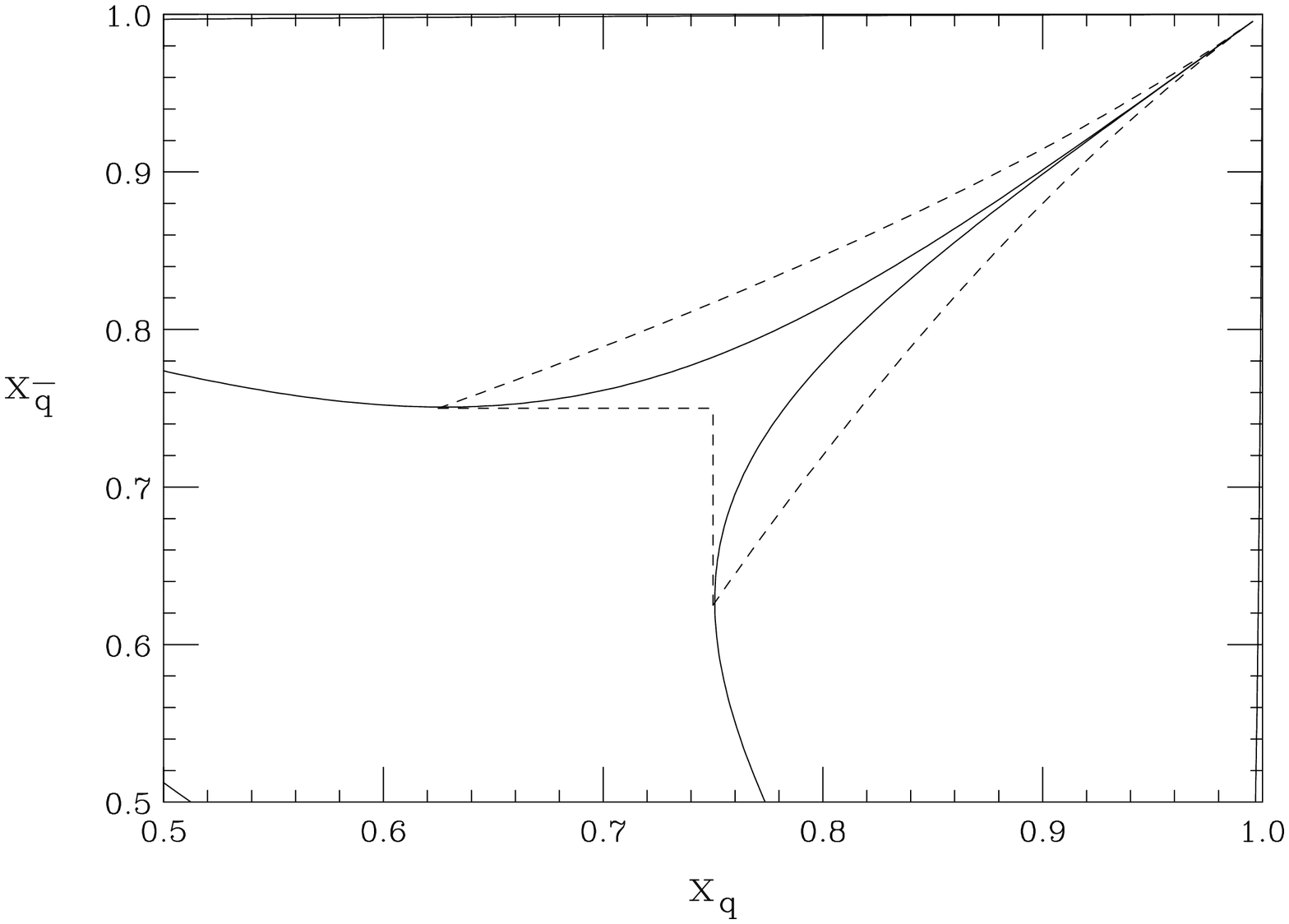,angle=90,width=15cm}
\end{center}
\caption{The soft region, with jet boundaries (solid) and mapped region
(dashed), for $m=5$ GeV, $Q^2=m_Z^2$.}
\label{fig_soft_map}
\end{figure}

Although the integral in eq.~(\ref{eq_dsigdead}) is finite, the integrand
diverges as one approaches the soft limit $x_q=x_\qb=1$ via the narrow
``neck'' of the dead region in fig.~\ref{fig_bfragn} or \ref{fig_bfragn_max}.
This could cause problems in generating $q\qb g$ configurations in the
dead region in order to apply a matrix element
correction~\cite{Seymour:1994we}.  To avoid such problems, one can
map the region $x_q,x_\qb > \frac 34$ into a region whose width vanishes
quadratically as $x_q, x_\qb\to 1$, as illustrated in fig.~\ref{fig_soft_map}.
The mapping shown is
\beeq
x_q \to x'_q &=& 1-\left[\frac 14 -(1-x_q)\right] =\frac 74 -x_q\;,
\nonumber\\
x_\qb \to x'_\qb &=& 1-2(1-x'_q)\left[\frac 34 -(1-x_q)\right]
= \frac 58 +\frac 12 x_q +\frac 32 x_\qb -2x_q x_\qb
\eeeq
when $x_q > x_\qb > \frac 34$.  Within the mapped region, the integrand
then has an extra weight factor of $2(1-x'_q)$ which regularizes the soft
divergence. When $x_\qb > x_q > \frac 34$, $x_q$ and  $x_\qb$ are
interchanged in both the mapping and the weight.

\section{Initial-initial colour connection}
Here we consider the inverse process $b+c\to a$ where $a$ is a colour singlet
of invariant mass $Q$ and $b,c$ are beam jets.  The kinematics are simple
because we take beam jets to be massless: in the c.m.\ frame
\beq
p_a = Q(1,\0t,0)\;,\;\;\; p_b= \half Q(1,\0t,1)
\;,\;\;\; p_c= \half Q(1,\0t,-1)\;.
\eeq
For emission of a gluon $g$ from $b$ we write
\beq\label{eq_iiqi}
q_i = \alpha_i p_b + \beta_i p_c + q_{\perp i}
\eeq
where $\qtr{g}=\kt$, $\qtr{a}=-\kt$, $\qtr{b}=\qtr{c}=\0t$.  Notice that
in this case the recoil transverse momentum is taken by the colour singlet
$a$ so we cannot preserve its 4-momentum.  We choose to preserve its mass
and rapidity, so that
\beq
\alpha_a = \beta_a =\sqrt{1+\kappa}\;,
\eeq
where as before $\kappa\equiv\kt^2/Q^2$.  Now we have
\beeq
\beta_b  &=& \alpha_c = 0\;,\;\;\; \alpha_g\beta_g = \kappa\;,\nonumber\\
\alpha_a &=& \alpha_b - \alpha_g\;,\;\;\;
\beta_a = \beta_c - \beta_g\;,
\eeeq
and our new variables in this case are
\beq\label{eq_iztk}
z=1-\frac{\alpha_g}{\alpha_b}\;,\;\;\;
\tk \equiv \frac{\tq^2}{Q^2} = \frac{\kappa}{(1-z)^2}\;.
\eeq
Thus we find
\beeq\label{eq_iialphas}
\alpha_a &=& \beta_a =\sqrt{1+(1-z)^2\tk}\;,\nonumber\\
\alpha_b &=& \frac{1}{z}\sqrt{1+(1-z)^2\tk}\;,\nonumber\\
\beta_c &=& \frac{1+(1-z)\tk}{\sqrt{1+(1-z)^2\tk}}\;.
\eeeq

\subsection{Phase space variables}
It is convenient to express the kinematics in terms of the ``reduced''
Mandelstam invariants:
\beq
\sb=(q_b+q_c)^2/Q^2\;,\;\;\; \tb=(q_b-q_g)^2/Q^2\;,\;\;\;
\ub=(q_c-q_g)^2/Q^2\;.
\eeq
The phase space limits are
\beq
1 < \sb < S/Q^2\;,\;\;\; 1-\sb < \tb < 0\;,\;\;\; \ub = 1-\sb-\tb
\eeq
where $S$ is the beam-beam c.m.\ energy squared.
In terms of the shower variables for beam jet $b$, we have
\beq
\sb = \alpha_b\beta_c =  \frac{1}{z}[1+(1-z)\tk]\;,\;\;\;
\tb = -\alpha_b\beta_g = -(1-z)\tk\;,\;\;\; \ub = -(1-z)\sb\;.
\eeq
Thus curves of constant $\tk$ in the $(\sb,\tb)$ plane are given by
\beq
\tb = \frac{\tk(1-\sb)}{\tk+\sb}
\eeq
and the Jacobian factor for conversion of the shower variables to the
Mandelstam invariants is
\beq
\frac{\partial(\sb,\tb)}{\partial(z,\tk)} = \frac{1-z}{z}\sb\;.
\eeq

\begin{figure}
\begin{center}
\epsfig{figure=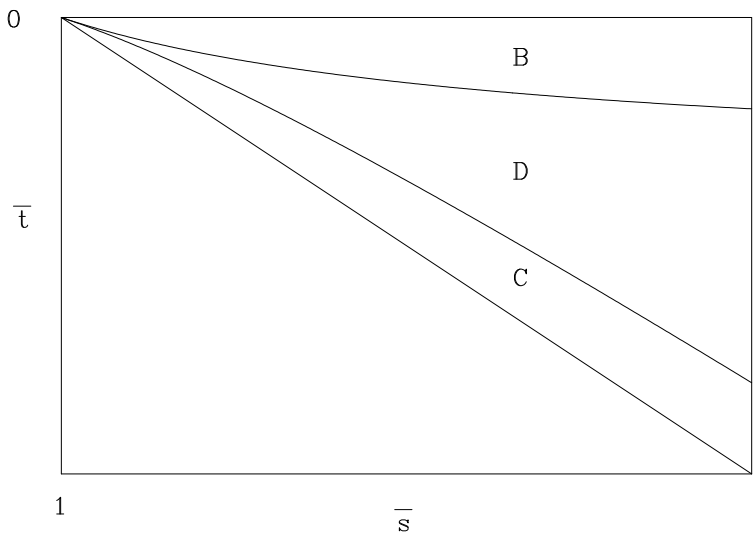,angle=90,width=10cm}
\end{center}
\caption{Beam jets (B,C) and dead region (D) in initial-state branching.}
\label{fig_DYps}
\end{figure}

For the other beam jet $c$ we have $\tb\leftrightarrow\ub$ and thus
\beq
\tb = \frac{\sb(1-\sb)}{\tk+\sb}\;.
\eeq
We see that in order for the jet regions to touch without overlapping
in the soft limit $\sb\to 1$, $\tb\to 0$, we need $\tk <\tk_b$ in jet
$b$ and  $\tk <\tk_c$ in jet $c$, where $\tk_c = 1/\tk_b$.  The most
symmetrical choice is $\tk_c =\tk_b =1$, as shown in fig.~\ref{fig_DYps},
but we can take $\tk_b$ or $\tk_c$ as large as we like.

\subsection{Example: Drell-Yan process}
Consider radiation from the quark in the Drell-Yan process,
$q\bar q\to g Z^0$.  In the laboratory frame we have
\beq
q_q = (Px_1,\0t,Px_1)\;,\;\;\;
q_{\qb} = (Px_2,\0t,-Px_2)
\eeq
where $P =\half\sqrt S$ is the beam momentum.  If we generated
the initial hard process  $q\bar q\to Z^0$ with momentum fractions
$x_q$, $x_\qb$ and we want to preserve the mass and
rapidity of the $Z^0$ we require
\beq
x_1 = x_q\alpha_b\;,\;\;\; x_2 = x_\qb\beta_c
\eeq
where $\alpha_b$ and $\beta_c$ are given by eqs~(\ref{eq_iialphas}).

The branching probability in the parton shower approximation is
\beq\label{eq_Pzk}
\frac{d^2 P}{dz\,d\tk}
= C_F\frac{\as}{2\pi}\,\frac{1}{\tk}\,\frac{1+z^2}{1-z}\;,
\eeq
which gives a differential cross section ($s=\sb Q^2$, etc.)
\beq\label{eq_dsigDYps}
\frac{1}{\sigma_0}\frac{d^2\sigma}{ds\,dt}=
\frac{D(x_1)D(x_2)}{D(x_q)D(x_\qb)}
\frac{\as}{2\pi}C_F\frac{s+u}{s^3 t u}
\left[s^2+(s+u)^2\right]
\eeq
where $\sigma_0$ is the Born cross section.
The functions $D(x_1)$ etc.\ are parton distribution functions in the
incoming hadrons; these factors take account of the change of
kinematics $x_q,x_\qb \to x_1,x_2$ discussed above.

The exact differential cross section for $q\bar q\to g Z^0$ to order $\as$ is
\beq\label{eq_dsigDYme}
\frac{1}{\sigma_0}\frac{d^2\sigma}{ds\,dt}=
\frac{D(x_1)D(x_2)}{D(x_q)D(x_\qb)}
\frac{\as}{2\pi}C_F\frac{Q^2}{s^3 t u}
\left[(s+t)^2+(s+u)^2\right]\;.
\eeq
Since $Q^2=s+t+u$ and $t\leq 0$, we see that the parton shower
approximation (\ref{eq_dsigDYps}) overestimates the exact expression,
becoming exact in the collinear or soft limit $t\to 0$.  Therefore the
gluon distribution in the jet regions can be corrected efficiently
by the rejection method, and the dead region can be filled using
the matrix element, as was done in \cite{Corcella:1999gs}.
The benefit of the new variables is that the angular distribution
of soft gluon emission requires no correction, provided the jet regions
touch without overlapping in the soft region.  As shown above, this will
be the case if the upper limits on $\tk$ satisfy $\tk_\qb = 1/\tk_q$.

\section{Initial-final colour connection}
Consider the process $a+b\to c$ where $a$ is a colour singlet and the
beam parton $b$ and outgoing parton $c$ are colour-connected.
An example is deep inelastic scattering,
where $a$ is a (charged or neutral) virtual gauge boson.
We need to preserve the 4-momentum of $a$ and therefore we work
in the Breit frame:
\beq
p_a = Q(0,\0t,-1)\;,\;\;\; p_b= \half Q(1+c,\0t,1+c)
\;,\;\;\; p_c= \half Q(1+c,\0t,-1+c)\;,
\eeq
where $p_a^2=-Q^2$, $p_b^2=0$, and $m_c^2=cQ^2$.  Notice that
the beam parton $b$ is always taken to be massless, but the
outgoing parton $c$ can be massive (e.g.\ in $W^+ d\to c$).

\subsection{Initial-state branching}\label{sec_ini_fin}
For emission of a gluon $g$ from the incoming parton $b$ we write
\beq
q_i = \alpha_i p_b + \beta_i n + q_{\perp i}
\eeq
where $\qtr{g}=\kt$, $\qtr{b}=\0t$, $\qtr{c}=-\kt$ and we choose
\beq\label{eq_ndef}
n = \half Q(1+c,\0t,-1-c)\;.
\eeq
To preserve $p_a=q_c+q_g-q_b$ we now require
\beq
\alpha_b-\alpha_c-\alpha_g = \beta_c+\beta_g-\beta_b= \frac{1}{1+c}
\eeq
whereas the mass-shell condition is
\beq
\alpha_i\beta_i\,Q^2(1+c)^2 = \qtr{i}^2 + q_i^2
\eeq
which gives
\beq\label{eq_iibetas}
1+c =\frac{c}{\alpha_c}+\kappa\left(\frac{1}{\alpha_c}+
\frac{1}{\alpha_g}\right)\;.
\eeq

The new variables for emission from the beam jet are as in
eq.~(\ref{eq_iztk}).  Substituting in (\ref{eq_iibetas}), we find
\beeq\label{eq_inifin_abs}
\alpha_b &=& \frac{1}{2z(1+c)}\left(1+c+(1-z)\tk +
\sqrt{[1+c+(1-z)\tk]^2-4z(1-z)\tk}\right)\;,\nonumber\\
\alpha_c &=& z\alpha_b - \frac{1}{1+c}\;,\;\;\;
\alpha_g = (1-z)\alpha_b\;,\;\;\;\beta_b=0\;,\nonumber\\
\beta_c &=& \frac{1}{1+c}\cdot
\frac{c+(1-z)^2\tk}{z(1+c)\alpha_b-1}\;,\;\;\;
\beta_g = \frac{(1-z)\tk}{(1+c)^2\alpha_b}\;.
\eeeq

\subsection{Final-state branching}\label{sec_fin_ini}
Next consider emission from the outgoing parton $c$. In this case we write
\beq
q_i = \alpha_i p_c + \beta_i p_b + q_{\perp i}
\eeq
To preserve $p_a=q_c+q_g-q_b$ we require
\beq
\alpha_c+\alpha_g-\alpha_b = \beta_b-\beta_c-\beta_g= 1
\eeq
whereas the mass-shell condition is now
\beq
\alpha_i\beta_i\,(Q^2+m_c^2) = \qtr{i}^2 + q_i^2 -\alpha_i^2 m_c^2\;.
\eeq
The new variables for emission from an outgoing parton are as in
eqs.~(\ref{eq_fztk},\ref{eq_fkt}) with $b$ replaced by $c$:
\beq\label{eq_fztkc}
z = \frac{\alpha_c}{\alpha_c+\alpha_g}\;,\;\;\;
\tk \equiv \frac{\tq^2}{Q^2}=\frac{1}{z^2}\left[
c+\frac{\kappa}{(1-z)^2}\right]\;.
\eeq
Thus in this case we find
\beeq\label{eq_ifalphas}
\alpha_b &=& 0\;,\;\;\; \alpha_c = z\;,\;\;\; \alpha_g = 1-z\;,\nonumber\\
\beta_b &=& \frac{1}{1+c}[1+c+z(1-z)\tk]\;,\nonumber\\
\beta_c &=& \frac{1-z}{1+c}[2c+z(1-z)\tk]\;,\nonumber\\
\beta_g &=& \frac{1-z}{1+c}[z^2\tk -2c]\;.
\eeeq

\subsection{Phase space variables}\label{sec_inifin_var}
In this process the invariant phase space variables are usually taken to be
\beq
x_p = \frac{Q^2}{2p_a\cdot q_b}\;,\;\;\;
z_p = \frac{q_c\cdot q_b}{p_a\cdot q_b}\;.
\eeq
In terms of the new variables for emission from the beam parton, we have
\beeq
x_p &=& \frac{1}{(1+c)\alpha_b} = 2z
\left(1+c+(1-z)\tk +\sqrt{[1+c+(1-z)\tk]^2-4z(1-z)\tk}\right)^{-1} \\
z_p &=& (1+c)\beta_c = \frac 12
\left(1-c-(1-z)\tk +\sqrt{[1+c+(1-z)\tk]^2-4z(1-z)\tk}\right)\;,
\eeeq
with the Jacobian
\beq\label{eq_inifin_jac_ini}
\frac{\partial(x_p,z_p)}{\partial(z,\tk)} = \frac{1}{\tk}
\left(\frac{1}{x_p}+\frac{1+c}{1-z_p}-2\right)^{-1}\;.
\eeq
In the soft limit $z=1-\eps$ we therefore find for the beam jet
\beq\label{eq_inifin_sof_ini}
x_p \sim \frac{1}{1+c}\left[1-\eps -\frac{\eps c\tk}{(1+c)^2}\right]
\;,\;\;\; z_p \sim 1-\frac{\eps \tk}{1+c}
\eeq
and
\beq\label{eq_inifin_jac_sof}
\frac{\partial(x_p,z_p)}{\partial(z,\tk)} \sim \frac{\eps}{(1+c)^2}\;.
\eeq

\begin{figure}
\begin{center}
\epsfig{figure=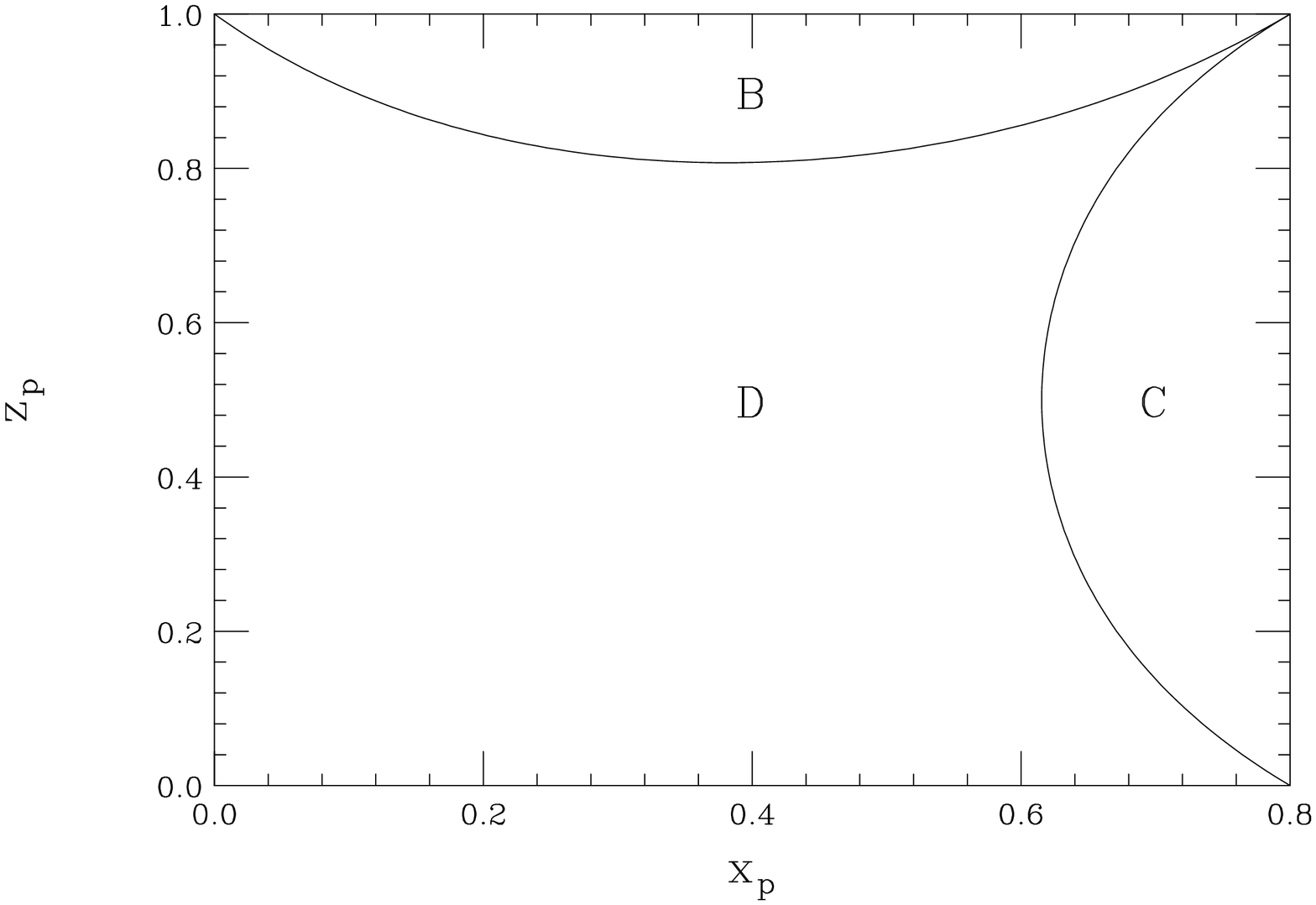,angle=90,width=10cm}
\end{center}
\caption{Beam jet (B), outgoing jet (C) and dead region (D) in initial-final
state branching: $c=0.25$, $\tk_b=1.25$, $\tk_c=1.5$.}
\label{fig_inifin_1}
\end{figure}

\begin{figure}
\begin{center}
\epsfig{figure=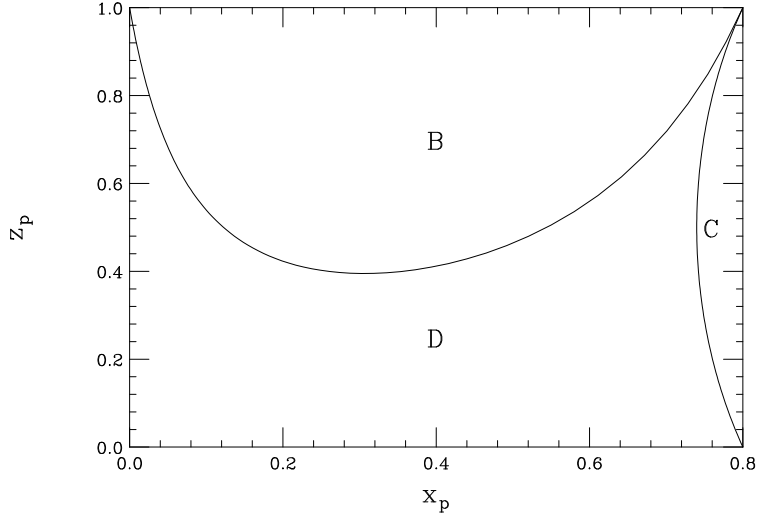,angle=90,width=10cm}
\end{center}
\caption{Beam jet (B), outgoing jet (C) and dead region (D) in initial-final
state branching: $c=0.25$, $\tk_b=10$, $\tk_c=0.40625$.}
\label{fig_inifin_10}
\end{figure}

In terms of the variables for emission from the outgoing parton,
\beq
x_p =\frac{1}{(1+c)\beta_b} = \frac{1}{1+c+z(1-z)\tk}\;,\;\;\;
z_p = \alpha_c = z\;,
\eeq
so the Jacobian is simply
\beq\label{eq_inifin_jac_fin}
\frac{\partial(x_p,z_p)}{\partial(z,\tk)} = z(1-z)x_p^2\;,
\eeq
and in the soft limit
\beq\label{eq_inifin_sof_fin}
x_p \sim  \frac{1}{1+c}\left[1-\frac{\eps \tk}{1+c}\right]\;,\;\;\;
z_p \sim 1-\eps\;,
\eeq
with the Jacobian again given by eq.~(\ref{eq_inifin_jac_sof}).
For full coverage of phase space in the soft limit we require
$\tk<\tk_b$ in jet $b$ and $\tk<\tk_c$ in jet $c$, where
\beq\label{eq_inifin_kbc}
\tk_b (\tk_c -c) = (1+c)^2\;.
\eeq
Thus the most symmetrical choice is $\tk_b=1+c$, $\tk_c=1+2c$, as shown in
fig.~\ref{fig_inifin_1}.  On the other hand, any larger or smaller
combination satisfying eq.~(\ref{eq_inifin_kbc}) is allowed, as
illustrated in fig.~\ref{fig_inifin_10} for $\tk_b=10$.

\subsection{Example: deep inelastic scattering}
Consider deep inelastic scattering on a hadron of momentum
$P^\mu$ by exchange of a virtual photon of momentum $q^\mu$.
If the contribution to the Born cross section from scattering
on a quark of momentum fraction $x_B=Q^2/2P\cdot q$ is
represented by $\sigma_0$ (a function of $x_B$ and $Q^2$),
then the correction due to single gluon emission is given by
\beq\label{eq_dsig_dis}
\frac{1}{\sigma_0}\frac{d^2\sigma}{dx_p\,dz_p}=
\frac{C_F\as}{2\pi}\frac{D(x_B/x_p)}{D(x_B)}
\frac{1+(x_p+z_p-1)^2}{x_p(1-x_p)(1-z_p)}\;.
\eeq

In the soft limit $x_p,z_p\to 1$ we have, from
eqs.~(\ref{eq_inifin_sof_ini},\ref{eq_inifin_sof_fin}) with $c=0$,
\beq
(1-x_p)(1-z_p)\sim \eps^2\tk
\eeq
and so
\beq\label{eq_dsig_dis_sof}
\frac{1}{\sigma_0}\frac{d^2\sigma}{dx_p\,dz_p}\sim
\frac{C_F\as}{\pi}\frac{1}{\eps^2\tk}\;,
\eeq
whereas the parton shower approximation gives
\beq\label{eq_dsig_dis_ps}
\frac{1}{\sigma_0}\frac{d^2\sigma}{dz\,d\tk}\sim
\frac{C_F\as}{\pi}\frac{1}{\eps\tk}\;.
\eeq
Since the Jacobian factor (\ref{eq_inifin_jac_ini}) or
(\ref{eq_inifin_jac_fin}) in this limit is simply $\eps$, the
shower approximation is exact in the soft limit.

\subsection{Example: $q\qb \to t\bar t$}
We denote the momenta in this process by $p_a+p_b \to p_c+p_d$
and the $2\to 2$ invariants by
\beq
\sb =  2p_a\cdot p_b\;,\;\;\;
\tb = -2p_a\cdot p_c\;,\;\;\;
\ub = -2p_a\cdot p_d\;,
\eeq
so that $\sb+\tb+\ub =0$.
Colour flows from $q$ to $t$ and anticolour from $\qb$ to $\bar t$.
Therefore the momentum transfer $q=p_a-p_c = p_d-p_b$ is carried
by a colour singlet and we preserve this 4-momentum during
showering.

For emission from the incoming light quark or the outgoing top 
quark, we work in the Breit frame for this system, where
\beq\label{eq_qqtt_q}
q = Q(0,\0t,1)\;,\;\;\; p_a= \half Q(1+c,\0t,1+c)
\;,\;\;\; p_c= \half Q(1+c,\0t,-1+c)
\eeq
with $Q^2=-\tb-m_t^2$ and $c=m_t^2/Q^2$. Then the treatment
of sects.~\ref{sec_ini_fin} and \ref{sec_fin_ini} can be applied
directly, with the substitution $b\to a$ since the emitting
system is now $(a,c)$ rather than $(b,c)$.  However, the
phase space variables are no longer those of sect.~\ref{sec_inifin_var}
since they involve the momenta of the $\qb$ and $\bar t$,
which in the frame (\ref{eq_qqtt_q}) take the general form
\beeq
p_b &=& [\half Q\sqrt{(1+c)^2+4K},\Qt,-\half Q(1+c)]\;,\nonumber\\
p_d &=& [\half Q\sqrt{(1+c)^2+4K},\Qt,\half Q(1-c)]\;,
\eeeq
where $K=\Qt^2/Q^2$ is related to the $2\to 2$ invariants:
\beq
\sb = \half Q^2 (1+c)^2\left[1+\sqrt{1+\frac{4K}{(1+c)^2}}\right]\;,\;\;\;
\tb = -Q^2 (1+c) \;,\;\;\; \ub = -\sb-\tb\;,
\eeq
and so
\beq
\Qt^2 = -\sb\left[1+\frac{\sb}{(1+c)\tb}\right]\;.
\eeq

For emission from the incoming light quark we define as in
sect.~\ref{sec_ini_fin}
\beq
q_i = \alpha_i p_a + \beta_i n + q_{\perp i}
\eeq
for $i=a,c,g$, where $\qtr{a}=\0t$, $\qtr{c}=-\kt$, $\qtr{g}=\kt$,
and $n$ is as in eq.~(\ref{eq_ndef}).
Then the $\alpha_i$'s and $\beta_i$'s are given by
eqs.~(\ref{eq_inifin_abs}) with the substitution $b\to a$. The
light antiquark and the antitop are not affected and therefore
$q_b=p_b$, $q_d=p_d$. This allows the complete kinematics of
the $2\to 3$ process to be reconstructed.   The $2\to 3$
invariants can be defined as in ref.~\cite{Frixione:2003ei}:
\beq
s=2q_a\cdot q_b\;,\;\;\;
t_1 = -2q_a\cdot q_c\;,\;\;\;
t_2 = -2q_b\cdot q_d\;,\;\;\;
u_1 = -2q_a\cdot q_d\;,\;\;\;
u_2 = -2q_b\cdot q_c\;.
\eeq
It is convenient to express $n=p_c-cq$ so that (for $i=a,c,g$)
\beq
q_i = (\alpha_i -c\beta_i)p_a + (1+c)\beta_i p_c + q_{\perp i}\;.
\eeq
Then we find
\beq\label{eq_qqtt_ini}
s = \alpha_a\sb\;,\;\;\;
t_1 = \alpha_a\beta_c(1+c)\tb\;,\;\;\;
t_2 = \tb\;,\;\;\;
u_1 = \alpha_a\ub\;,\;\;\;
u_2 = \beta_c(\ub-c\tb)-\alpha_c\sb -2 \kt\cdot\Qt\;.
\eeq 

For emission from the outgoing top we use the results of
sect.~\ref{sec_fin_ini}, again with the substitution $b\to a$.
Thus we now have for $i=a,c,g$
\beq
q_i = \alpha_i p_c + \beta_i p_a + q_{\perp i}
\eeq
where the $\alpha_i$'s and $\beta_i$'s are given by
eqs.~(\ref{eq_ifalphas}) with $b\to a$, and we find that
\beq\label{eq_qqtt_fin}
s = \beta_a\sb\;,\;\;\;
t_1 = \alpha_c\beta_a\tb\;,\;\;\;
t_2 = \tb\;,\;\;\;
u_1 = \beta_a\ub\;,\;\;\;
u_2 = \alpha_c\ub-\beta_c\sb -2 \kt\cdot\Qt\;.
\eeq

Similar formulae to eqs.~(\ref{eq_qqtt_ini}) and (\ref{eq_qqtt_fin}),
with the replacements $a\to b$ and $c\to d$, will hold for the case of
gluon emission from the colour-connected $(\qb\bar t)$ system.
Using these relations, one can study the distribution of gluon
radiation in the parton shower approximation and compare it with the
exact $q\qb\to t\bar t g$ matrix element. Agreement will be good in
the soft and/or collinear regions but there will be regions of hard,
wide-angle gluon emission in which matrix element corrections should
be applied.  Alternatively, the above equations can be used to formulate
a modified subtraction scheme for combining fixed-order and parton-shower
results, as was done in ref.~\cite{Frixione:2003ei} for a different
parton-shower algorithm.

\section{Decay colour connection}
Consider the process $b\to ca$ where $a$ is a colour singlet and the
decaying parton $b$ and outgoing parton $c$ are colour-connected.
Examples are bottom quark decay, $b\to cW^*$, and top decay, $t\to bW$.
Here we have to preserve the  4-momentum of the decaying parton $b$
and therefore we work in its rest frame,
\beq\label{eq_dec_pbcg}
p_b = m_b(1,\0t,0)\;,\;\;\; p_c= \half m_b(1-a+c,\0t,\lambda)
\;,\;\;\; p_a= \half m_b(1+a-c,\0t,-\lambda)\;,
\eeq
where $a=m_a^2/m_b^2$, $c=m_c^2/m_b^2$ and now
\beq
\lambda = \lambda(1,a,c) = \sqrt{(1+a-c)^2-4a} = \sqrt{(1-a+c)^2-4c}\;.
\eeq

\subsection{Initial-state branching}
For emission of a gluon $g$ from the decaying parton $b$ we write
\beq\label{eq_decqi}
q_i = \alpha_i p_b + \beta_i n + q_{\perp i}
\eeq
where $\qtr{g}=\kt$, $\qtr{c}=-\kt$, $\qtr{b}=\0t$ and we choose
\beq
n = \half m_b(1,\0t,1)\;,
\eeq
i.e.\ aligned along $p_c$ in the rest frame of $b$. The mass-shell
conditions give
\beq
\beta_a = \frac{a}{\alpha_a}-\alpha_a\;,\;\;\;\;
\beta_c = \frac{c+\kappa}{\alpha_c}-\alpha_c\;,\;\;\;\;
\beta_g = \frac{\kappa}{\alpha_g}-\alpha_g\;,
\eeq
with $\kappa=\kt^2/m_b^2$. From momentum conservation
\beq\label{eq_mom}
\alpha_a + \alpha_c + \alpha_g = 
\frac{a}{\alpha_a}+\frac{c+\kappa}{\alpha_c}+\frac{\kappa}{\alpha_g} =1\;.
\eeq
Recall that in initial-state branching of a heavy object
our new evolution variable is given by eq.~(\ref{eq_dec_tqi}),
so we have
\beq
\alpha_g=1-z\;,\;\;\;\; \kappa = (\tk-1)(1-z)^2
\eeq
where $\tk = \tq^2/m_b^2 >1$.  Introducing for brevity the notation
\beq\label{eq_wuv}
w = 1-(1-z)(\tk-1)\;,\;\;\;
u = 1+a-c-(1-z)\tk\;,\;\;\;
v = \sqrt{u^2-4awz}\;,
\eeq
from eq.~(\ref{eq_mom}) we find
\beq\label{eq_alphaac}
\alpha_a = \frac{u+v}{2w}\;,\;\;\;
\alpha_c = 1-\alpha_a -\alpha_g = z -\alpha_a\;.
\eeq

\subsection{Final-state branching}
For radiation from the outgoing parton $c$ we write
\beq
q_i = \alpha_i p_c + \beta_i n + q_{\perp i}
\eeq
where $p_c$ is given by eq.~(\ref{eq_dec_pbcg}).
Since the colour-connected parton $b$ is at rest in our working frame
of reference, the choice of the light-like vector $n$ in this case is
somewhat arbitrary.  By analogy with the cases treated earlier, we
choose it to be opposite to that used for the radiation from $b$,
i.e.\ along the direction of the colour singlet $a$: 
\beq
n = \half m_b(\lambda,\0t,-\lambda)\;.
\eeq
The kinematics are then identical with those for final-final connection
(sect.~\ref{sec_finfin}), with the replacement $b\to c$, $c\to a$.

\subsection{Phase space variables}
As in sect.~\ref{sec_finfin}, it is convenient to use the
Dalitz plot variables, which in this case are 
\beq
x_i = \frac{2 q_i\cdot p_b}{m_b^2}\;.
\eeq
For emission from the decaying parton $b$ we have
$x_i = 2\alpha_i +\beta_i$ and hence, from eq.~(\ref{eq_alphaac}),
\beq\label{eq_xacg}
x_a = \frac{u+v}{2w} + \frac{u-v}{2z}\;,\;\;\;
x_c = w+z-x_a\;,\;\;\; x_g = 2-w-z = (1-z)\tk\;,
\eeq
with the Jacobian factor
\beq
\frac{\partial(x_a,x_g)}{\partial(z,\tk)} =
(1-z)\left[\frac{u+v}{2w^2}-\frac{u-v}{2z^2}+\frac{a(w-z)^2}{vwz}\right]\;.
\eeq
 
In the soft limit $z\to 1-\eps$ we find
\beq\label{eq_dec_xag}
x_a\sim 1+a-c-\eps\tk'_b\;,\;\;\;
x_g\sim \eps\tk_b
\eeq
where
\beq\label{eq_dec_tkp}
\tk'_b = \lambda+\frac{\tk_b}{2}(1-a+c-\lambda)\;.
\eeq

For emission from the outgoing parton $c$ we have, from eq.~(\ref{eq_xcbg})
with the replacement $b\to c$, $c\to a$:
\beeq\label{eq_decfin}
x_a &=& 1+a-c- z(1-z)\tk \nonumber \\
x_c &=& (2-x_a)r + (z-r)\sqrt{x_a^2-4a}\\
x_g &=& (2-x_a)(1-r) - (z-r)\sqrt{x_a^2-4a}\nonumber
\eeeq
where
\beq\label{eq_decr}
r = \frac 12\left(1+\frac{c}{1+a-x_a}\right)\;.
\eeq
In the soft limit we have from eq.~(\ref{eq_soft})
\beq
x_a\sim 1+a-c-\eps\tk_c\;,\;\;\;
x_g \sim \eps\tk'_c
\eeq
where
\beq
\tk'_c = \lambda+\frac{\tk_c}{2c}(1-a+c-\lambda)\;.
\eeq
For full coverage of the soft region we require
\beq
\frac{\tk_b}{\tk'_b} = \frac{\tk'_c}{\tk_c}
\eeq
which gives in this case
\beq\label{eq_tkdec}
(\tk_b -1)(\tk_c -c) =  \frac 14 (1-a+c+\lambda)^2\;.
\eeq
Note that, while there is no upper limit on $\tk_b$, the largest value
that can be chosen for $\tk_c$ is given by the equivalent of
eq.~(\ref{eq_fftkmax}),
\beq\label{eq_dectkmax}
\tk_c < 4(1+a-2\sqrt{c} -c)\;.
\eeq

\subsection{Example: top decay}
In the decay $t\to Wbg$ we have $a=m_W^2/m_t^2=0.213$ and
$c=m_b^2/m_t^2=0.026$, so for simplicity we neglect $c$.
Then for radiation from the top we have from eq.~(\ref{eq_xacg})
\beq
x_W = \frac{u+v}{2w} + \frac{u-v}{2z}\;,\;\;\;
x_g = (1-z)\tk\;,
\eeq
where $u,v,w$ are given by eqs.~(\ref{eq_wuv}) with $c\to 0$.
The phase space is the region
\beq
0 <  x_g < 1-a \;,\;\;\;\;
1-x_g+\frac{a}{1-x_g} < x_W < 1+a\;.
\eeq
Notice that for real $x_W$ we require $u^2>4awz$, i.e.
\beq
1<\tk<1 + a\left[1-\sqrt{\frac{z(1-a)}{a(1-z)}}\right]^2\;,
\eeq
and
\beq
1- \frac{1-a}{\tk+2\sqrt{a(\tk-1)}} < z < 1\;.
\eeq
Thus there is no upper limit on $\tk$, but the range of $z$ becomes
more limited as $\tk$ increases.

\begin{figure}
\begin{center}
\epsfig{figure=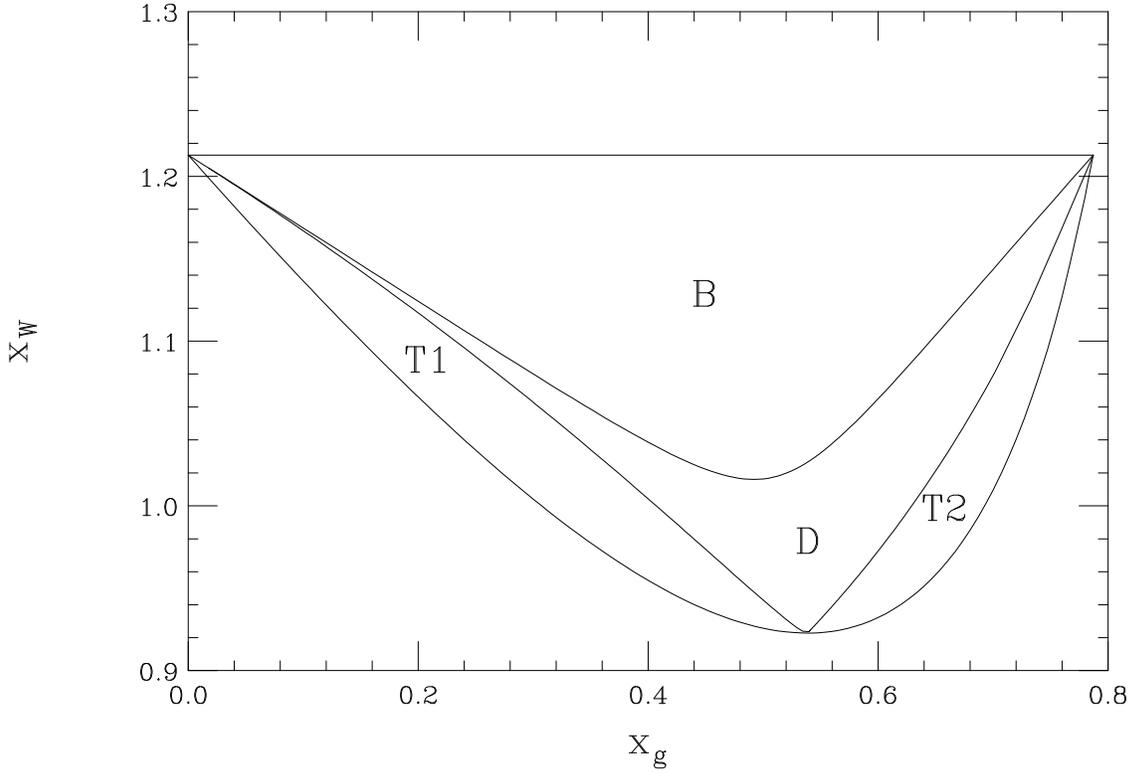,angle=90,width=15cm}
\end{center}
\caption{Phase space for decay $t\to Wbg$, with symmetric choice of
emission regions for the $b$ (B) and the $t$ (T1,T2),
and the dead region (D).}
\label{fig_tdec_sym}
\end{figure}

For radiation from the $b$ we have from eq.~(\ref{eq_decfin})
\beeq\label{eq_topb}
x_W &=& 1+a- z(1-z)\tk \nonumber \\
x_g &=& \half(2-x_W) - (z-\half)\sqrt{x_W^2-4a}\;.
\eeeq
To cover the soft region we require $\tk<\tk_t$ for emission from
the top quark and $\tk<\tk_b$ for that from the bottom, where
eq.~(\ref{eq_tkdec}) gives
\beq\label{eq_tkb_tkt}
\tk_b = \frac{(1-a)^2}{\tk_t-1}\;.
\eeq
The most symmetrical choice would therefore appear to be
$\tk_b =\tk_t-1 = 1-a = 0.787$, as illustrated in fig.~\ref{fig_tdec_sym}.

\begin{figure}
\begin{center}
\epsfig{figure=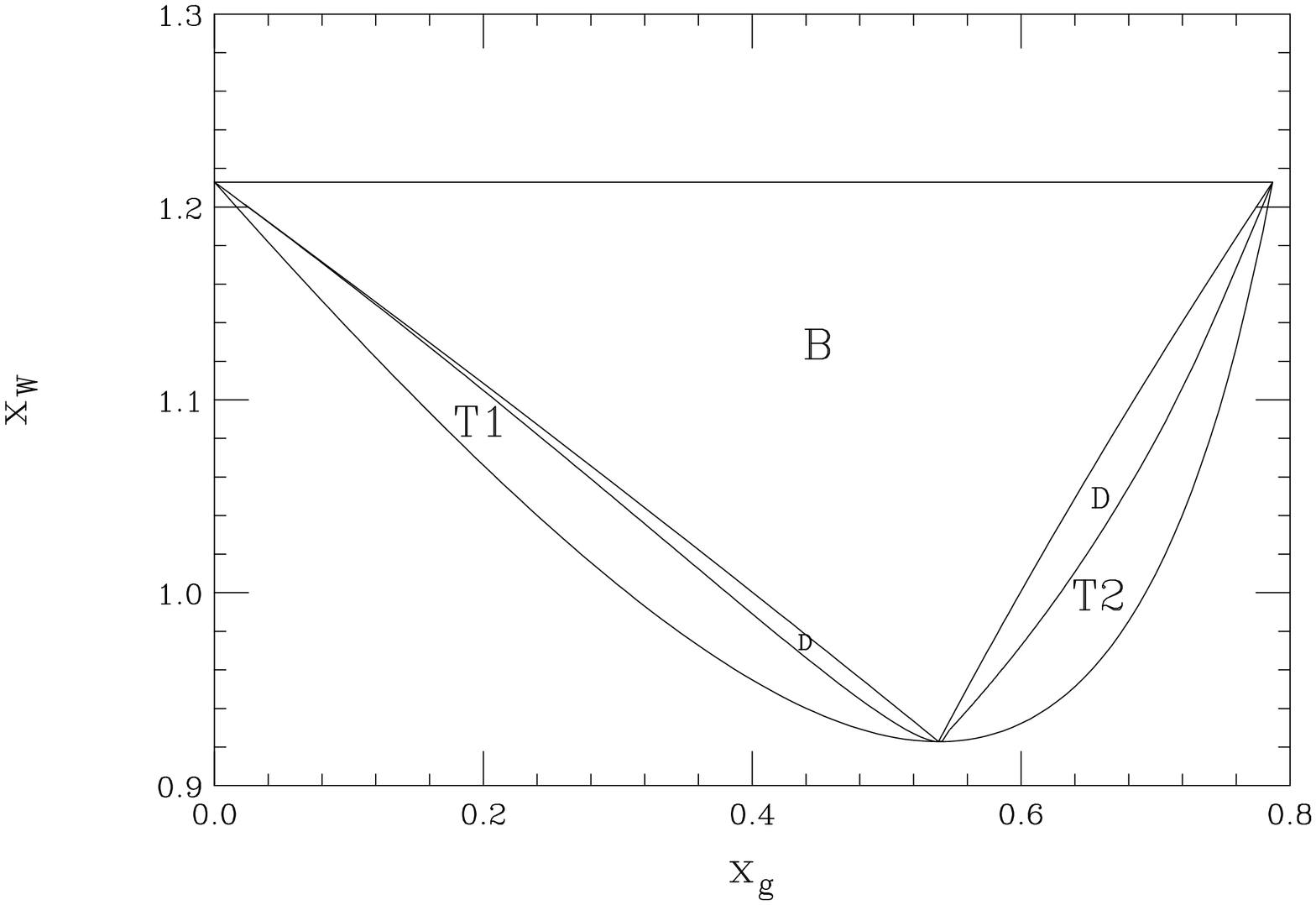,angle=90,width=15cm}
\end{center}
\caption{Phase space for decay $t\to Wbg$, with maximal region (B)
for emission from the $b$, together with complementary regions of
emission from the $t$ (T1,T2) and the dead region (D).}
\label{fig_tdec_brad}
\end{figure}

As mentioned above, there is no upper limit on $\tk_t$.  Thus the region
covered by gluon emission from the top quark can be as large as we like.
However, eq.~(\ref{eq_dectkmax}) tells us that the upper limit
for radiation from the $b$ is
\beq\label{eq_tkbmax}
\tk_b < 4 (1-\sqrt a)^2 = 1.16\;,
\eeq
and correspondingly $\tk_t > 1+\frac 14 (1+\sqrt a)^2 = 1.53$.
Figure \ref{fig_tdec_brad} shows this maximal region that can be covered
by emission from the $b$, together with the complementary regions of
emission from the $t$.

We note from figs.~\ref{fig_tdec_sym} and \ref{fig_tdec_brad} that,
for any value of $\tk_t$, the region for emission from the top quark
consists of two distinct parts that touch at the point
$x_g=1-\sqrt a$, $x_W=2\sqrt a$, where the $W$ boson is at rest:
a subregion T1 which includes the soft limit $x_g\to 0$ and a hard
gluon region T2.

%\subsubsection{Exact matrix element}
The exact $t\to Wbg$ differential decay rate to first order in $\as$
is given in \cite{Corcella:1998rs}:
\beeq
\frac{1}{\Gamma_0}\frac{d^2\Gamma}{dx_W dx_g} &=& \frac{\as}{\pi}\frac{C_F}
{(1+a-x_W)x_g^2}\Biggl\{x_g-\frac{(1+a-x_W)(1-x_g)+x_g^2}{1-a}\nonumber\\
&&+x_g\frac{(x_W+x_g-1-a)^2}{2(1-a)^2}
+\frac{2a(1+a-x_W)x_g^2}{(1-a)^2(1+2a)}\Biggr\}
\eeeq
where $\Gamma_0$ is the lowest-order decay rate.  In the soft region
$x_W\to 1+a$, $x_g\to 0$ this becomes
\beq
\frac{1}{\Gamma_0}\frac{d^2\Gamma}{dx_W dx_g}\sim \frac{\as}{\pi}\frac{C_F}
{x_g}\left[\frac{1}{1+a-x_W}-\frac{1}{(1-a)x_g}\right]\;.
\eeq

For soft gluon emission ($1-z=\eps\to 0$) from the top quark we have from
eqs.~(\ref{eq_dec_xag},\ref{eq_dec_tkp})
\beq
x_W\sim 1+a-\eps(1-a)\;,\;\;\;
x_g\sim \eps\tk
\eeq
and so the exact form of the soft gluon distribution is
\beq
\frac{1}{\Gamma_0}\frac{d^2 \Gamma}{dx_g dx_W} \sim \frac{\alpha_s}{\pi}
\frac{C_F}{\eps^2} f_t(\tk)
\eeq
where
\beq
f_t(\tk)= \frac{\tk-1}{(1-a)\tk^2}\;.
\eeq
In the same region the parton shower approximation (\ref{eq_dPqqg}) gives
\beq
\frac{d^2 P}{dx_g dx_W} \sim \frac{1}{(1-a)\eps}\frac{d^2 P}{dz\,d\tk}
\sim \frac{\alpha_s}{\pi}\frac{C_F}{(1-a)\epsilon^2\tk}\left(1-
\frac{1}{\tk}\right)=\frac{\alpha_s}{\pi}\frac{C_F}{\eps^2}f(\tk)\;.
\eeq
Thus we see that, for emission from the top quark,
the shower approximation is exact in the soft limit.  At higher
gluon energies, inside the region T1 the parton shower
overestimates the exact matrix element and can therefore be
corrected easily by the rejection method.  In the hard gluon region
T2, which contributes only a small finite correction to the cross
section, the parton shower overestimates the matrix element
at lower values of $x_g$ but underestimates it at the highest values.
Therefore a combination of rejection and matrix element correction
is needed in this region.

For emission from the bottom quark in the soft limit, we use the results
of sect.~\ref{sec_finfin_sof} with the substitution $b\to 0$, $c\to a$
to obtain
\beq
x_W\sim 1+a-\eps\tk\;,\;\;\;
x_g\sim \eps\left(1-a+\frac{\tk}{1-a}\right)\;.
\eeq
Therefore the exact soft gluon distribution in the $b$ jet should be
\beq
\frac{1}{\Gamma_0}\frac{d^2 \Gamma}{dx_g dx_W} \sim \frac{\alpha_s}{\pi}
\frac{C_F}{\eps^2} f_b(\tk)
\eeq
where
\beq
f_b(\tk)= \frac{1}{(1-a)\tk}\left[1+\frac{\tk}{(1-a)^2}\right]^{-2}\;.
\eeq
On the other hand the parton shower approximation in this case gives simply
\beq
\frac{d^2 P}{dx_g dx_W} \sim \frac{1}{(1-a)\eps}\frac{d^2 P}{dz\,d\tk}
\sim \frac{\alpha_s}{\pi}\frac{C_F}{(1-a)\epsilon^2\tk}\;.
\eeq
Thus the soft gluon distribution in the $b$ jet region is overestimated
by a factor of
\beq
\left[1+\frac{\tk}{(1-a)^2}\right]^2\;,
\eeq
which can be corrected by the rejection method.  This factor varies from
1 to 5.2 for the symmetric choice of the $b$ jet region $\tk_b=1-a$
depicted in fig.~\ref{fig_tdec_sym}. For the maximal  $b$ jet shown
in fig.~\ref{fig_tdec_brad}, it rises to 8.3.  Since the shower
approximation is exact in the soft limit for emission from the top,
one can reduce the amount of soft correction required by decreasing the
$b$ jet region and increasing that for top emission,
in accordance with eq.~(\ref{eq_tkb_tkt}).  However, for large
values of $\tk_t$ the dead region moves near to the collinear
singularity at $x_W=1+a$ and a large hard matrix element correction
becomes necessary.
  
\section{Conclusions}
We have presented a new formulation of the parton-shower approximation
to QCD matrix elements, which offers a number of advantages
over previous ones.  Direct angular ordering of the shower ensures
a good emulation of important QCD coherence effects, while the
connection between the shower variables and the Sudakov-like
representation of momenta (\ref{eq_qi}) simplifies the
kinematics and their relation to phase space invariants.
The use of mass-dependent splitting functions with the
new variables allows an accurate description of soft gluon
emission from heavy quarks over a wide angular region, including
the collinear direction. The separation of showering into
contributions from pairs of colour-connected hard partons
permits a general treatment of coherence effects, which should be
reliable at least to leading order in the number of colours.
Since the formulation is slightly different for initial-
and final-state showering, we have provided formulae for
all colour-connected combinations of incoming and outgoing
partons.

As mentioned in the Introduction, this new shower formulation is a key
element of the event generator \HWP\ \cite{Gieseke:2003_20}, and detailed
results of its implementation for $e^+e^-$ annihilation will be presented
shortly~\cite{Gieseke:2003_19}.

\section*{Acknowledgments}
We are most grateful to the other \HWP\ authors, Mike Seymour and Alberto
Ribon, for helpful discussions.  S.G.\ also acknowledges fruitful
conversations with Frank Krauss. S.G.\ and P.S.\ thank CERN Theory Division
for hospitality during part of this work.  This research was supported
by the UK Particle Physics and Astronomy Research Council.

\end{document}